\documentclass[12pt]{article}
\pdfoutput=1

\usepackage{putex}
\usepackage{graphicx}
\usepackage{caption}
\usepackage{amsmath}
\usepackage{array}
\usepackage{subcaption}
\usepackage{epstopdf}
\usepackage{enumerate}
\usepackage{cite}
\usepackage{youngtab}
\usepackage{tensor}
\usepackage{slashed}
\usepackage[aligntableaux=center]{ytableau}
\usepackage[utf8]{inputenc}
\usepackage{rotating}
\usepackage{bigfoot}
\usepackage[
      colorlinks=true,
      linkcolor=blue,
      urlcolor=blue,
      filecolor=black,
      citecolor=red,
      ]{hyperref}

\newcommand {\be} {\begin {equation}}
\newcommand {\ee} {\end {equation}}

\newcommand {\no} {\nonumber}
\newcommand {\bes} {\begin {equation*}}
\newcommand {\ees} {\end {equation*}}

\newsavebox{\numberedequationbox}
\newcommand{\es}[2]{%
\begin{equation}\label{#1}%
\sbox{\numberedequationbox}{$\displaystyle\begin{aligned}#2\end{aligned}$}%
\ifdim\wd\numberedequationbox>.90\linewidth
\resizebox{.90\linewidth}{!}{\usebox{\numberedequationbox}}%
\else\usebox{\numberedequationbox}\fi
\end{equation}}
\def\g{\gamma}

\newcommand{\Z}{\mathbb{Z}}

\newcommand{\cA}{{\mathcal A}}

\newcommand{\cG}{{\mathcal G}}

\newcommand{\cO}{{\mathcal O}}

\newcommand{\cS}{{\mathcal S}}

\newcommand{\cM}{{\mathcal M}}

\newcommand{\beq}{\begin{equation}}
\newcommand{\eeq}{\end{equation}}

\def\ie{\begin{equation}\begin{aligned}}
\def\fe{\end{aligned}\end{equation}}

\numberwithin{equation}{section}

\newcommand\Tstrut{\rule{0pt}{2.3ex}}       % "top" strut
\newcommand\Bstrut{\rule[-1.3ex]{0pt}{0pt}} % "bottom" strut
\newcommand\TBstrut{\Tstrut\Bstrut}         % "top and bottom" strut

\def\<{\langle}
\def\>{\rangle}

\begin{document}

\preprint{}

\institution{oxford}{Mathematical Institute, University of Oxford,
Woodstock Road, Oxford, OX2 6GG, UK}
\institution{Exile}{Department of Particle Physics and Astrophysics, Weizmann Institute of Science, Rehovot, Israel}

\title{ABJM at Strong Coupling from M-theory, Localization, and Lorentzian Inversion}

\authors{Luis F. Alday\worksat{\oxford}, Shai M. Chester\worksat{\Exile}, and Himanshu Raj\worksat{\Exile}}

\abstract{
We study the stress tensor multiplet four-point function in the 3d maximally supersymmetric ABJ(M) theory with Chern-Simons level $k=2$, which in the large $N$ limit is holographically dual to weakly coupled M-theory on $AdS_4\times S^7/\mathbb{Z}_2$. We use the Lorentzian inversion to compute the 1-loop correction to this holographic correlator coming from Witten diagrams with supergravity $R$ and the first higher derivative correction $R^4$ vertices, up to a finite number of contact terms that contribute to low spins where the inversion formula does not converge. We find a precise match with the corresponding terms in the 11d M-theory S-matrix by taking the flat space limit, which is not sensitive to these contact terms. We then conjecturally fix these contact terms by analytically continuing the inversion formula below its expected range of convergence, and verify this conjecture using supersymmetric localization. Finally, we compare some of the 1-loop CFT data to non-perturbative in $N$ bounds from the numerical conformal bootstrap, which we compute at unprecedently high accuracy, and find that the 1-loop corrections saturate the bounds in the large $N$ regime, which extends the previously observed match at tree level.
}
\date{}

\maketitle

\tableofcontents

\section{Introduction}
\label{intro}

Holographic correlators are correlation functions of single trace operators in CFTs with holographic duals, such that the large $N$ limit of the CFT correlators is dual to scattering of gravitons or KK modes in weakly coupled AdS gravity. One reason to study these correlators is to learn about quantum gravity on curved spacetime. Another motivation is that there is a concrete limit \cite{Penedones:2010ue,Polchinski:1999ry,Susskind:1998vk,Giddings:1999jq,Fitzpatrick:2011hu,Fitzpatrick:2011jn} of the AdS correlator that gives the flat space S-matrix, so we can use CFT to study quantum gravity in flat space. For CFTs with M-theory duals, which is the subject of this paper, this motivation is especially compelling, because the M-theory S-matrix cannot be computed from flat space methods even in principle beyond the lowest few terms in the small Planck length expansion \cite{Green:1997as, Russo:1997mk, Green:2005ba}. Instead, \cite{Chester:2018aca} proposed that the M-theory S-matrix could be defined to all orders by first computing the dual CFT correlator at large $N$ and then taking the flat space limit.

Holographic correlators were originally computed at large $N$ using Witten diagrams derived from the explicit AdS supergravity action, see e.g. \cite{Arutyunov:2000py,Arutyunov:2002ff,Arutyunov:2002fh,DHoker:1999bve,DHoker:1999mqo,DHoker:1999kzh}. Recently, a more powerful analytic bootstrap approach was introduced in \cite{Rastelli:2017udc} that computes tree level holographic correlators in the leading supergravity approximation purely based on crossing symmetry, analyticity, superconformal symmetry, and the flat space limit. This approach was generalized to higher derivative corrections to tree level supergravity in \cite{Alday:2014tsa,Goncalves:2014ffa} following the general discussion in \cite{Heemskerk:2009pn}, where the coefficients of these corrections are no longer completely determined by symmetry, but can be fixed using theory-specific inputs like supersymmetric localization \cite{Chester:2018aca,Binder:2018yvd,Binder:2019jwn,Binder:2019mpb,Chester:2019jas,Chester:2020dja,Chester:2020vyz,Chester:2019pvm} for 3d and 4d CFTs or protected sectors \cite{Chester:2018dga} for 6d CFTs. 

The cutting edge of the large $N$ analytic bootstrap is 1-loop, which can be computed up to a finite number of contact terms by ``squaring'' the tree level CFT data of all double-trace operators in the correlator \cite{Aharony:2016dwx}.  More precisely, we apply crossing symmetry to this data to compute the double-discontinuity, which according to the Lorentzian inversion formula \cite{Caron-Huot:2017vep} determines the entire correlator up to a finite number of contact terms discussed in \cite{Heemskerk:2009pn} that contribute to CFT data of low spins.\footnote{At finite $N$ the Lorentzian inversion formula in fact converges for CFT data of all spins \cite{Caron-Huot:2017vep,Caron-Huot:2018kta,Lemos:2021azv}, but this convergence gets worse in the large $N$ expansion.} This program was carried out for 1-loop diagrams involving the supergravity $R$ and first higher derivative correction $R^4$ vertices for 4d $SU(N)$ $\mathcal{N}=4$ SYM \cite{Alday:2017xua,Alday:2018pdi,Alday:2018kkw,Alday:2019nin,Alday:2017vkk,Aprile:2017bgs,Aprile:2017qoy,Aprile:2019rep,Drummond:2019hel,Drummond:2020dwr,Aprile:2020luw}, which is dual to Type IIB string theory on $AdS_5\times S^5$, and 6d $(2,0)$ theory \cite{Alday:2020tgi}, which is dual to $AdS_7\times S^4$ for the $A_{N-1}$ theories and $AdS_7\times S^4/\mathbb{Z}_2$ for the $D_N$ theories.\footnote{For CFTs dual to higher spin gravity, 1-loop terms were also computed in 3d in \cite{Aharony:2018npf}, see also \cite{Binder:2021cif,Turiaci:2018nua,Li:2019twz,Silva:2021ece} for tree level results.} The $R|R$ diagram has a single contact term ambiguity that was fixed using localization for $\mathcal{N}=4$ SYM in \cite{Chester:2019pvm}. However, higher loops diagrams such as $R|R^4$ have too many ambiguities to be fixed using localization, and for 6d $(2,0)$ theory there are no localization results available. These ambiguities also cannot be fixed by comparing to the known S-matrix in the flat space limit, which was used to fix tree level higher derivative terms in \cite{Goncalves:2014ffa,Binder:2019jwn,Chester:2020dja,Chester:2019jas,Chester:2020vyz}, since the contact term ambiguities are purely AdS features that disappear in the flat space limit.

In this paper we will extend the 1-loop analytic bootstrap to 3d maximally supersymmetric ABJ(M) theory \cite{Aharony:2008ug,Aharony:2008gk} with Chern-Simons level $k=2$, which is holographically dual to M-theory on $AdS_4\times S^7/\mathbb{Z}_2$, and propose how to fix 1-loop contact term ambiguities using an analytic continuation of the Lorentzian inversion formula. There are two such maximally supersymmetric CFTs with gauge groups $U(N)_2\times U(N)_{-2}$ or $U(N+1)_2\times U(N)_{-2}$, which are called ABJM or ABJ respectively, but they are identical when expanded at large central charge\footnote{The central charge is defined in \eqref{stress} as the coefficient of the canonically-normalized stress-tensor two point function, which has been calculated to all orders in $1/N$ through supersymmetric localization \cite{Chester:2014fya} using the results of \cite{Jafferis:2010un} and \cite{Closset:2012vg}.} $c_T\sim N^{\frac32}$,\footnote{This is because from the M-theory point of view, the two theories differ by a torsion flux, i.e.~a discrete holonomy of the 3-form field on a torsion 3-cycle of $S^7 / \Z_2$ \cite{Aharony:2008gk}.  This torsion flux affects the CFT data only through non-perturbative effects.} so we will refer to both theories collectively as ABJ(M). We will study the stress tensor multiplet correlator $\langle 2222\rangle$, where $\langle pppp\rangle$ denotes the correlator of the bottom component of the $p$th lowest single trace half-BPS multiplet, which are dual to the corresponding $p$th lowest scalar KK mode in the dimensional reduction of M-theory on AdS. We will find it convenient to work with the Mellin transform $M(s,t;\sigma,\tau)$ of $\langle2222\rangle$, where $\sigma,\tau$ parameterize the $R$-symmetry dependence and $s,t$ are Mellin variables that are related to the 11d Mandelstam variables in the flat space limit. The large $c_T$ expansion of $M$ is constrained by the analytic bootstrap to take the form
\es{M2222}{
M(s,t;\sigma,\tau)&=c_T^{-1}M^R+c_T^{-\frac53}B^{R^4}_4M^4+c_T^{-2}(M^{R|R}+B^{R|R}_4M^4)\\
&+c_T^{-\frac73}(B^{D^6R^4}_4M^4+B^{D^6R^4}_6M^6+B^{D^6R^4}_7M^7)\\
&+c_T^{-\frac{23}{9}}(B^{D^8R^4}_4M^4+B^{D^8R^4}_6M^6+B^{D^8R^4}_7M^7+B^{D^8R^4}_8M^8)\\
&+c_T^{-\frac83}(M^{R|R^4}+B^{R|R^4}_4M^4+B^{R|R^4}_6M^6+B^{R|R^4}_7M^7+B^{R|R^4}_8M^8)+\dots\,,
}
where the $M$'s are functions of $s,t,\sigma,\tau$ with numerical coefficients $B$ that can depend on $k$. The tree level terms\footnote{Since $c_T$ is the only expansion parameter, we can only distinguish between tree and loop terms at low orders where they have different power of $c_T$.} at orders $c_T^{-1}$, $c_T^{-\frac53}$, and $c_T^{-\frac73}$ were previously computed for both $k=1,2$ ABJ(M) in \cite{Zhou:2017zaw}, \cite{Chester:2018aca}, and  \cite{Binder:2018yvd}, respectively, while this paper will focus on the 1-loop terms $R|R$ at order $c_T^{-2}$ and $R|R^4$ at order $c_T^{-\frac83}$ for $k=2$ ABJ(M).\footnote{We will also show some results for the $R^4|R^4$ term at order $c_T^{-{10}/{3}}$, but we will not consider the $R|D^6R^4$ term that contributes at the same order} 

As in the 4d and 6d cases, double trace long operators are degenerate in the generalized free field theory (GFFT) at $c_T\to\infty$ limit, so their tree level CFT data at orders $c_T^{-1}$ for $R$ and $c_T^{-\frac53}$ for $R^4$ must be unmixed to get the 1-loop corrections we consider. For $k=2$ ABJ(M), this unmixing requires the average of GFFT OPE coefficients obtained from $\langle pppp\rangle$ for even $p$, as well as the average of $c_T^{-1}$ and $c_T^{-\frac53}$ anomalous dimensions obtained from $\langle 22pp\rangle$ for even $p$. For $k=1$ ABJM, the 1-loop double discontinuity would also receive contributions from the OPE coefficients of double trace long operators with odd twists, which generically contribute to the 1-loop double discontinuity of large $N$ 3d CFTs \cite{Aharony:2018npf}. These degenerate contributions must be similarly unmixed using GFFT OPE coefficients in $\langle pppp\rangle$ for odd $p$ as well as the average of $c_T^{-1}$ and $c_T^{-\frac53}$ OPE coefficients from $\langle 22pp\rangle$ for odd $p$. These odd $p$ terms do not contribute to the $k=2$ theory, because they are projected out by the orbifold.\footnote{In the 4d and 6d cases, all double trace multiplets have even twists, so only anomalous dimensions contribute to the 1-loop double discontinuity. The only difference between the orbifold and the non-orbifold cases is that we restrict the sum over tree level anomalous dimensions to even $p$ for the orbifold cases.} We computed the average GFFT OPE coefficients from $\langle pppp\rangle$ for general $p$ by computing the full superconformal block expansion up to $p=9$ using the superconformal Ward identities in \cite{Dolan:2004mu} and then guessing the general $p$ formula, similar to what we did in \cite{Alday:2020tgi} for the 6d case. For the tree $\langle 22pp\rangle$ data for even $p$, we extracted the average anomalous dimension from the $R$ correlator given in \cite{Alday:2020dtb} as well as the $R^4$ correlator that we compute here using the known M-theory term in the flat space limit. The odd $p$ case is technically much harder, because the tree level supergravity $\langle 22pp\rangle$ cannot be written in terms of a finite number of $\bar D_{r_1,r_2,r_3,r_4}(U,V)$ functions, unlike the even $p$ case in 3d or the general $p$ case in 4d and 6d.\footnote{In 4d and 6d, the correlator can be written in terms of just a few $\bar D_{r_1,r_2,r_3,r_4}(U,V)$ with integer arguments for any $p$. In 3d and even $p$, the number of $\bar D_{r_1,r_2,r_3,r_4}(U,V)$ grows with $p$ and has negative and half-integer arguments, as we will show in the main text.} We will thus only discuss $k=2$ ABJ(M) in this paper, where we use the even $p$ data to compute $R|R$, $R|R^4$ and $R^4|R^4$ using the Lorentzian inversion formula up to the finite number of contact terms discussed above, extract the low-lying CFT data for spins unaffected by these contact terms, and successfully compare to the M-theory S-matrix terms computed in \cite{Russo:1997mk,Green:1997as,Alday:2020tgi} in the flat space limit, which is unaffected by these contact terms.

We then analytically continue the Lorentzian inversion formula to extract CFT data of spins that are affected by the contact term ambiguities. For $R|R$, we find that this analytic continuation works for all CFT data, and in particular allows us to fix the contact term $B^{R|R}_4M^4$ in \eqref{M2222} to zero, where $M^{R|R}$ is defined so that its CFT data is analytic in spin for all values. For $R|R^4$, we find that the analytic continuation of the inversion formula works for all CFT data except that affected by the $B^{R|R^4}_4M^4$ contact term, which allows us to fix the other three contact terms in \eqref{M2222}. We then apply two localization constraints from \cite{Chester:2018aca,Agmon:2017xes} and \cite{Binder:2018yvd,Binder:2019mpb} to further constrain the amplitude. For $R|R$ we find that one of the localization constraints independently fixes $B^{R|R}_4=0$,\footnote{We are only able to write $M^{R|R}$ up to a finite number of polynomial in $s,t$ ambiguities that are in principle fixed by superconformal symmetry, but are difficult to fix in practice. As such we could not impose the localization constraint in \cite{Binder:2018yvd,Binder:2019mpb} that requires us to integrate $M^{R|R}$, but we were still able to impose the localization constraint from \cite{Chester:2018aca,Agmon:2017xes} that simply fixes certain protected OPE coefficients, which can be extracted from $R|R$ using the Lorentzian inversion formula without knowing the explicit Mellin amplitude as we will discuss in the main text. For $M^{R|R^4}$, we managed to compute the complete Mellin amplitude, so we could impose both localization constriants.} which confirms the result from the conjectured analytic continuation, while for $R|R^4$ we use one constraint to fix the remaining $B^{R|R^4}_4$ coefficient and the second constraint as a nontrivial check.

Finally, we compare the CFT data extracted from the 1-loop Mellin amplitudes to the numerical bootstrap. In \cite{Agmon:2017xes}, it was conjectured that the $k=2$ ABJ(M) theory saturates the numerical bootstrap bounds for $\langle2222\rangle$, which was motivated by comparing the bounds to the all orders in $1/c_T$ calculation of short operator OPE coefficients computed from supersymmetric localization \cite{Agmon:2017xes,Nosaka:2015iiw,Dedushenko:2016jxl,Kapustin:2009kz}. This conjecture was further checked in \cite{Chester:2018lbz}, where the $c_T^{-1}$ correction to all CFT data in $\langle2222\rangle$ was found to saturate the bootstrap bounds at sufficiently large $c_T$, but this correction does not depend on $k$. To compare higher order in $1/c_T$ corrections that depend on $k$, one must keep in mind that the large $c_T$ expansion is asymptotic, so either one must look at very high values of $c_T$, which require extremely high numerical bootstrap accuracy, or focus on terms in the $1/c_T$ expansion where the asymptotic expansion still converges (i.e. subsequent terms have decreasing coefficients). We focus on the $c_T^{-2}$ terms in the OPE coefficients of semishort operators, which are the most converged corrections beyond $O(c_T^{-1})$, and find that these corrections noticeably improve the saturation of the bootstrap bounds, which we compute at much higher accuracy than the previous studies \cite{Chester:2014fya,Chester:2014mea,Agmon:2017xes,Agmon:2019imm}. We do not have enough accuracy yet to extract sufficient CFT data to fix the $D^8R^4$ term in \eqref{M2222}, which would fulfill the goal of \cite{Chester:2018aca} of deriving new terms in the M-theory S-matrix from CFT. Nevertheless, this is the first successful comparison of analytic $c_T^{-2}$ terms to numerical bootstrap, which improves the many tree level checks performed in various other contexts in \cite{Beem:2016wfs,Beem:2015aoa,Chester:2018lbz,Binder:2021cif,Binder:2020ckj}, and is a necessary step toward the goal of deriving $D^8R^4$.

 The rest of this paper is organized as follows. In Section~\ref{4points} we compute the explicit superblock decomposition of $\langle qqpp\rangle$ for $q\leq p$ and $p\leq9$. We use these superblock expansions to extract the average GFFT OPE coefficients and tree level anomalous dimensions that we will need to compute the 1-loop data for $k=2$ ABJM. In Section~\ref{1loop}, we use this data to compute the 1-loop corrections to the stress tensor correlator up to contact term ambiguities. We then match these 1-loop terms to the M-theory S-matrix in the flat space limit. We also extract CFT data from the 1-loop correlators using the Lorentzian inversion formula as well as a projection method. In Section \ref{1loopContact}, we fix the contact term ambiguities at orders $c_T^{-2}$ and $c_T^{-\frac83}$ by combining constraints from supersymmetric localization with an analytic continuation of the Lorentzian inversion formula.  In Section \ref{numerics} we compare some of these 1-loop analytic results to bounds from the numerical conformal bootstrap, which we compute at much higher accuracy than previous studies. Finally, in Section~\ref{conc} we discuss future directions.  Several technical details are given in various Appendices.  We also include an ancillary \texttt{Mathematica} notebook, which includes many of our more complicated explicit results.

\section{Four-point functions at large $c_T$}
\label{4points}

We start by discussing the large $c_T\sim N^{3/2}$ expansion of four-point functions of the dimension $\frac p2$ scalar bottom component of half-BPS supermultiplets in $\mathcal{N}=8$ ABJ(M) theory, and derive the data needed for the 1-loop terms for $k=2$ ABJ(M) in the following sections. First we will review the constraints of the 3d $\mathcal{N}=8$ superconformal algebra $\mathfrak{osp}(8|4)\supset\mathfrak{so}(5)\oplus\mathfrak{so}(8)_R$ on $\langle ppqq\rangle$ following \cite{Agmon:2019imm}, and we explicitly perform the superblock expansion for $p,q\leq9$. We then discuss the generalized free field theory (GFFT) that describes the $c_T\to\infty$ limit, which we use to compute average OPE coefficients of double-trace singlet long multiplets in $\langle qqpp\rangle$. Afterwards, we consider tree level corrections to $\langle 22pp\rangle$ for even $p$, which we use to derive the average anomalous dimension of singlet long multiplets at orders $1/c_T$ and $1/c_T^{5/3}$ that correspond to tree level supergravity and $R^4$, respectively. Finally, we discuss higher orders in the large $c_T$ expansion of $\langle2222\rangle$, which will be our main focus in the rest of the paper.

\subsection{Block expansion of $\langle qqpp\rangle$}
\label{qqpp}

We consider half-BPS superconformal primaries $S_p$ in 3d $\mathcal{N}=8$ SCFTs that are scalars with $\Delta=\frac p2$ and transform in the $[00p0]$ of $\mathfrak{so}(8)_R$,\footnote{The convention we use in defining these multiplets is that the supercharges transform in the ${\bf 8}_v = [1000]$ irrep of $\mathfrak{so}(8)_R$.} where $p=1,2,\dots$. The first such interacting operator is $S_2$, which is the bottom component of the stress tensor multiplet. We can denote these operators as traceless symmetric tensors $S_{I_1\dots I_p}( x)$ of $\mathfrak{so}(8)_R$, where $I_i=1,\dots8$. We can avoid indices by introducing an auxiliary polarization vector $Y^I$ that is constrained to be null, $Y_i\cdot Y_i=0$, and then define
\es{S}{
S_p( x,Y)\equiv S_{I_1\dots I_p}Y^{I_1}\cdots Y^{I_p}\,.
}

Consider the four-point functions $\langle qqpp\rangle$ of four $S_p(x,Y)$'s, where $q\leq p$. Conformal and $\mathfrak{so}(8)_R$ symmetry fixes these correlators to take the form
\es{4point}{
\langle S_q( x_1,Y_1)S_q( x_2,Y_2)S_p( x_3,Y_3)S_p( x_4,Y_4) \rangle=\frac{(Y_1\cdot Y_2)^q(Y_3\cdot Y_4)^p}{| x_{12}|^{q}| x_{34}|^{p}}\mathcal{G}_{qp}(U,V;\sigma,\tau)\,,
}
where we define
 \es{uvsigmatauDefs}{
  U \equiv \frac{{x}_{12}^2 {x}_{34}^2}{{x}_{13}^2 {x}_{24}^2} \,, \qquad
   V \equiv \frac{{x}_{14}^2 {x}_{23}^2}{{x}_{13}^2 {x}_{24}^2}  \,, \qquad
   \sigma\equiv\frac{(Y_1\cdot Y_3)(Y_2\cdot Y_4)}{(Y_1\cdot Y_2)(Y_3\cdot Y_4)}\,,\qquad \tau\equiv\frac{(Y_1\cdot Y_4)(Y_2\cdot Y_3)}{(Y_1\cdot Y_2)(Y_3\cdot Y_4)} \,,
 }
 with $x_{ij}\equiv x_i-x_j$. Since \eqref{4point} is a degree $q$ polynomial in each $Y_i$ separately, the quantity $\mathcal{G}_{qp}(U,V;\sigma,\tau)$ is a degree $q$ polynomial in $\sigma$ and $\tau$. We parametrize these polynomials in terms of eigenfunctions $Y_{[0\,m\,\,2n-2m\,0]}(\sigma, \tau)$ of the $\mathfrak{so}(8)_R$ quadratic Casimir for irreps $[0\,m\, 2n-2m\,0]$ that appear in the tensor product of $[00q0]\otimes[00q0]$, so that $n=0,1,\dots q$ and $m=0,\dots n$.  The polynomials $Y_{[0\,m\, 2n-2m\,0]}(\sigma, \tau)$ can be computed explicitly \cite{Nirschl:2004pa}, and we give the explicit values for $q=2,\dots 9$ in the attached \texttt{Mathematica} file. We then expand $\mathcal{G}_{qp}(U,V;\sigma,\tau)$ in terms of this basis as
 \es{Ybasis}{
\mathcal{G}_{qp}(U,V;\sigma,\tau)=\sum^{q}_{n=0}\sum_{m=0}^{n}Y_{[0\,m\, 2n-2m\,0]}(\sigma, \tau) \mathcal{G}_{qp}^{[0\,m\, 2n-2m\,0]}(U,V)\,,
}
and furthermore expand $\mathcal{G}_{qp}^{[0\,m\, 2n-2m\,0]}(U,V)$ in conformal blocks $G_{\Delta,\ell}(U,V)$ as
\es{blockExp}{
\mathcal{G}_{qp}^{[0\,m\, 2n-2m\,0]}(U,V)=\sum_{\Delta,\ell}\lambda_{qq\cO_{\Delta,\ell,[0\,m\, 2n-2m\,0]}}\lambda_{pp\cO_{\Delta,\ell,[0\,m\, 2n-2m\,0]}}G_{\Delta,\ell}(U,V)\,,
}
where $\cO_{\Delta,\ell,[0\,m\, 2n-2m\,0]}$ are conformal primaries with scaling dimension $\Delta$ and spin $\ell$ in irrep $[0\,m\, 2n-2m\,0]$ that appear in $S_{q}\times S_{q}$ with OPE coefficient $\lambda_{qq\cO_{\Delta,\ell,[0\,m\, 2n-2m\,0]}}$. The 3d conformal blocks were computed in various series expansions in \cite{Dolan:2011dv,Dolan:2003hv,Kos:2013tga}, which we review in our conventions in Appendix \ref{block3dApp}.

The correlator is further constrained by the superconformal Ward identities \cite{Dolan:2004mu}:
\es{ward}{
\left[z\partial_z -  \frac12\alpha \partial_\alpha\right] \mathcal{G}_{qp}(z,\bar{z};\alpha, \bar{\alpha})|_{\alpha = \frac1z} =
\left[\bar{z}\partial_{\bar{z}} -  \frac12{\bar\alpha} \partial_{\bar{\alpha}}\right] \mathcal{G}_{qp}(z,\bar{z};\alpha, \bar{\alpha})|_{\bar{\alpha}=\frac{1}{\bar{z}}} &= 0\,,
}
where $z,\bar z$ and $\alpha,\bar\alpha$ are written in terms of $U,V$ and $\sigma,\tau$, respectively, as
\es{UVtozzbar}{
U=z\bar z\,,\quad V=(1-z)(1-\bar z)\,,\qquad\qquad \sigma=\alpha\bar\alpha\,,\quad \tau=(1-\alpha)(1-\bar\alpha)\,.
}
We can satisfy these constraints by expanding $ \mathcal{G}_{qp}$ in superconformal blocks as
\es{SBDecomp}{
      \mathcal{G}_{qp}(U,V;\sigma,\tau)=\sum_{\mathcal{M}\in S_{q}\times S_{q}}\lambda_{qq\mathcal{M}} \lambda_{pp\mathcal{M}} \mathfrak{G}_{\mathcal{M}}(U,V;\sigma,\tau)\,,
}
where $\mathfrak{G}_{\mathcal{M}}$ are superblocks for each supermultiplet $\mathcal{M}$ that appears in $S_{q}\times S_{q}$ (and $S_{p}\times S_{p}$)  with OPE coefficients $\lambda_{qq\mathcal{M}}$ (and $\lambda_{pp\mathcal{M}}$). The multiplets that can appear in the OPE are \cite{Ferrara:2001uj,Agmon:2019imm}:
\es{opemultEq}{
&S_q\times S_q=\text{Id}+\sum_{n=1}^q{(B,+)}_{n,0}^{[0\,0\,2n\,0]}+\sum_{n=1}^q\sum_{m=2,4,\dots,n}{(B,2)}_{n,0}^{[0\,m\,2n-2m\,0]}\\
&+\sum_{\ell=0,2,\dots}\sum_{n=1}^{q-1}{(A,+)}_{n+1+\ell,\ell}^{[0\,0\,2n\,0]}+\sum_{n=1}^{q-1}\Big[\sum_{\ell=0,2,\dots}\sum_{m=2,4,\dots,n}{(A,2)}_{n+1+\ell,\ell}^{[0\,m\,2n-2m\,0]}+\sum_{\ell=1,3,\dots}\sum_{m=1,3,\dots,n}{(A,2)}_{n+1+\ell,\ell}^{[0\,m\,2n-2m\,0]}\Big]\\
&+\sum_{n=0}^{q-2}\Big[\sum_{\ell=0,2,\dots}\sum_{m=0,2,\dots,n}{(A,0)}_{\Delta>n+1+\ell,\ell}^{[0\,m\,2n-2m\,0]}+\sum_{\ell=1,3,\dots}\sum_{m=1,3,\dots,n}{(A,0)}_{\Delta>n+1+\ell,\ell}^{[0\,m\,2n-2m\,0]}\Big]\,,
}
where we denote superconformal multiplets other than the identity $\text{Id}$ by $X_{\Delta,\ell}^{[a_1\, a_2 \, a_3 \, a_4]}$, with $(\Delta,\ell)$ and $[a_1\, a_2 \, a_3 \, a_4]$ representing the $\mathfrak{so}(3, 2)$ and $\mathfrak{so}(8)_R$ quantum numbers of the superconformal primary, while $X$ denotes the type of shortening condition. The $(A,0)^{[0\,m\,2n-2m\,0]}_{\Delta> n+\ell+1,\ell}$ multiplets that appear here are unprotected. When $\Delta$ saturates the bound we in general get semishort multiplets $(A,2)^{[0\,m\,2n-2m\,0]}_{ n+\ell+1,\ell}$ or $(A,+)^{[0\,0\,2n\,0]}_{ n+\ell+1,\ell}$, which are $\frac14$ or $\frac18$ BPS, respectively. The only exception is $(A,0)^{[0000]}_{\Delta> \ell+1,\ell}$ whose unitarity bound gives conserved currents that cannot appear in the interacting theories we consider. Finally, the $(B,2)^{[0\,n-m\,2n\,0]}_{ n,0}$ and $(B,+)^{[0\,0\,2n\,0]}_{n,0}$ are short multiplets, where the former is $\frac14$ BPS, while the latter are the half-BPS multiplets whose bottom component we called $S_p$. The lowest such multiplet is always the stress tensor multiplet $(B,+)_{1,0}^{[0020]}$, whose OPE coefficient squared is fixed by the conformal Ward identity \cite{Osborn:1993cr} to be inversely proportional to the coefficient of the canonically normalized stress tensor two-point function:
\es{stress}{
  \langle T_{\mu\nu}(\vec{x}) T_{\rho \sigma}(0) \rangle = \frac{c_T}{64} \left(P_{\mu\rho} P_{\nu \sigma} + P_{\nu \rho} P_{\mu \sigma} - P_{\mu\nu} P_{\rho\sigma} \right) \frac{1}{16 \pi^2 \vec{x}^2} \,, \qquad P_{\mu\nu} \equiv \eta_{\mu\nu} \nabla^2 - \partial_\mu \partial_\nu \,,
}
where $c_T$ is normalized so that $c_T^\text{free}=16$ for the free $\mathcal{N}=8$ theory of eight massless real scalars and Majorana fermions. In this normalization we get the precise relationship
\es{cTolam}{
\lambda_{pp{(B,+)}_{1,0}^{[0020]}}=\frac{8p}{{c_T}}\,.
}
We will be mostly interested in $\langle2222\rangle$, whose multiplets are summarized in Table \ref{opemult}. Note that we introduce simpler notation for these multiplets, e.g. $(B,+)^{[0040]}_{2,0}\equiv (B,+)$, since not counting the stress tensor multiplet, only one operator of each type appears.

\begin{table}
\centering
\begin{tabular}{|c|c|r|c|c|}
\hline
Type    & $(\Delta,\ell)$     & $\mathfrak{so}(8)_R$ irrep  &spin $\ell$ & Name \\
\hline
$(B,+)$ &  $(2,0)$         & ${\bf 294}_c = [0040]$& $0$ & $(B, +)$ \\ 
$(B,2)$ &  $(2,0)$         & ${\bf 300} = [0200]$& $0$ & $(B, 2)$ \\
$(B,+)$ &  $(1,0)$         & ${\bf 35}_c = [0020]$ & $0$ & Stress \\
$(A,+)$ &  $(\ell+2,\ell)$       & ${\bf 35}_c = [0020]$ &even & $(A,+)_\ell$ \\
$(A,2)$ &  $(\ell+2,\ell)$       & ${\bf 28} = [0100]$ & odd & $(A,2)_\ell$ \\
$(A,0)$ &  $\Delta\ge \ell+1$ & ${\bf 1} = [0000]$ & even & $(A,0)_{\Delta,\ell}$\\
$\text{Id}$ &  $(0,0)$ & ${\bf 1} = [0000]$ & even & $\text{Id}$\\
\hline
\end{tabular}
\caption{The possible superconformal multiplets in the $\cO_\text{Stress} \times  \cO_\text{Stress}$ OPE\@.  The $\mathfrak{so}(3, 2) \oplus \mathfrak{so}(8)_R$ quantum numbers are those of the superconformal primary in each multiplet.}
\label{opemult}
\end{table} 

We then compare \eqref{SBDecomp} to \eqref{Ybasis} and \eqref{blockExp} to see that the superblocks are finite linear combinations of conformal blocks 
\es{GExpansion}{
\mathfrak{G}_{\mathcal{M}}=\sum^{q}_{n=0}\sum_{m=0}^{n}Y_{[0\,m\, 2n-2m\,0]}(\sigma, \tau)  \sum_{\cO\in\mathcal{M}} \frac{\lambda_{qq\cO_{\Delta,\ell,[0\,m\, 2n-2m\,0]}} \lambda_{pp\cO_{\Delta,\ell,[0\,m\, 2n-2m\,0]}}   }{ \lambda_{qq\mathcal{M}} \lambda_{pp\mathcal{M}}  }G_{\Delta,\ell}(U,V)\,,
}
where $\cO_{\Delta,\ell,[0\,m\, 2n-2m\,0]}$ are conformal primaries that appear in $\mathcal{M}$, which can be derived using the Racah-Speiser algorithm in \cite{Cordova:2016emh}. For $\langle qqpp\rangle$, we fixed all of the pairs of OPE coefficients $\lambda_{qq\cO_{\Delta,\ell,[0\,m\, 2n-2m\,0]}}\lambda_{pp\cO_{\Delta,\ell,[0\,m\, 2n-2m\,0]}}$ in terms of the single pair of OPE coefficients $\lambda_{qq\mathcal{M}} \lambda_{pp\mathcal{M}} $ in \eqref{SBDecomp} by applying the Ward identities to the small $z,\bar z$ expansion of the superblocks, where we used the small $z,\bar z$ expansion of the conformal blocks as given in \cite{Dolan:2003hv} and reviewed in Appendix \ref{block3dApp}. We give the results for the $s$-channel of $\langle qqpp\rangle$ for $q=2,\dots,9$ in the attached \texttt{Mathematica} file.

\subsection{Generalized free field theory at $c_T\to\infty$}
\label{strong}

We now use the superblocks computed in the previous section to perform the superblock expansion for the GFFT that describes the $c_T\to\infty$ limit of $\langle qqpp\rangle$ and $\langle pppp\rangle$ for $k=1,2$ ABJ(M) theory. Recall from the Introduction that in fact both $k=1,2$ ABJ(M) have the same half-BPS correlators at this order, except that all correlators involving $S_p$ for odd $p$ vanish for the $k=2$ theory. In particular, both theories are described by a GFFT where the operators $S_p$ are treated as generalized free fields with two point functions $\langle S_p(x_1,Y_1) S_q(x_2,Y_2)\rangle=\delta_{pq}\frac{(Y_1\cdot Y_2)^p}{|x_{12}|^{p}}$. We can then compute $\langle qqpp\rangle$ (for $q\leq p$) using Wick contractions to get
\es{Ninf}{
\cG_{qp}^{(0)}=1+\delta_{qp}\left(U^{\frac p2}\sigma^p+\frac{U^{\frac p2}}{V^{\frac p2}}\tau^p\right)\,,
}
which can be expanded in the superblocks of the previous section to extract OPE coefficients. If several operators have the same quantum numbers at this order, then we can only compute the average of their OPE coefficients. Such a degeneracy occurs for the double trace long multiplet $(A,0)^{[0000]}_{\Delta,\ell}$ operator $S_p\partial_{\mu_1}\dots\partial_{\mu_\ell}(\partial^2)^nS_p$ with spin $\ell$ and twist $ t\equiv \Delta-\ell=p+2n\geq2$. For $t\geq2$, there are $t-1$ such degenerate operators because of the different ways of adding $p$ and $n$ to get the same twist, which we label using the degeneracy label $I$. We denote the GFFT OPE coefficient of these operators in the $S_p\times S_p$ OPE by $\lambda^{(0)}_{p,t,\ell,I}$, and note that only twists $t=p,p+2,\dots$ appear in this OPE at leading order. By expanding in superblocks for $p=2,\dots, 9$ we found the general formula
\es{ppppLam}{
\langle (\lambda^{(0)}_{p,t,\ell})^2\rangle&=\frac{45 \sqrt{\pi } (2 \ell+1) 4^{p-2} (p-1) p \Gamma \left(\frac{t-1}{2}\right) \Gamma \left(\ell+\frac{t}{2}\right) \Gamma (\ell+t+3) \Gamma \left( \frac p2+\frac t2+2\right)
   \Gamma \left(  \ell+\frac p2+\frac t2+\frac52\right)}{\Gamma (p+2) \Gamma (p+4) \Gamma \left(\frac{t+6}{2}\right) \Gamma \left(\ell+t+\frac{5}{2}\right) \Gamma \left(
    \ell+\frac t2+\frac72\right) \Gamma \left(-\frac p2+\frac t2+1\right) \Gamma \left(  \ell-\frac p2+\frac t2+\frac 32\right)}\,,
}
which reproduces the $p=2,3$ values given in \cite{Chester:2014fya,Agmon:2019imm}. Note that the average $\langle \lambda^{(0)}_{q,t,\ell} \lambda^{(0)}_{p,t,\ell} \rangle$ trivially vanishes because $ \cG_{qp}=1$ for $q\neq p$ at GFFT.

\subsection{Tree level $\langle 22pp\rangle$}
\label{22pptree}

We now consider the $1/c_T$ and $1/c_T^{5/3}$ terms in $\langle22pp\rangle$, which corresponds to tree level supergravity $R$ and $R^4$ in the bulk description, respectively, and whose CFT data is needed to compute loops with these vertices in the following section. We expand $\cG_{2p}$ (which we will denote as $\cG_{p}$) in \eqref{4point} as well as the long multiplet CFT data to this order as
\es{Hlarge}{
\cG_{p}(U,V;\sigma,\tau)&=\cG^{(0)}_{p}+c_T^{-1}\cG^R_{p}+c_T^{-\frac53}\cG^{R^4}_{p}+\dots\,\\
\Delta_{t,\ell,I}&=t+\ell+c_T^{-1}\gamma^{R}_{t,\ell,I}+c_T^{-\frac53}\gamma^{R^4}_{t,\ell,I}+\dots\,,\\
(\lambda_{p,t,\ell,I})^2&=(\lambda^{(0)}_{p,t,\ell,I})^2+c_T^{-1}(\lambda^{R}_{p,t,\ell,I})^2+c_T^{-\frac53}(\lambda^{R^4}_{p,t,\ell,I})^2+\dots\,.\\
}
A similar expansion exists for the OPE coefficients of the protected operators, although of course their scaling dimensions are fixed. Using these expansions, we can write the superblock expansion for $\cG_{p}$ in \eqref{SBDecomp} at large $c_T$ as
\es{SGexp}{
&\cG_{p}^R(U,V)={128p} \mathfrak{G}_\text{Stress}(U,V;\sigma,\tau) +\hspace{-.2in}\sum_{\mathcal{M}_{\Delta,\ell}\in\{(B,+),(B,2),(A,2)_\ell,(A,+)_\ell\}}\hspace{-.2in}\lambda^R_{22\mathcal{M}} \lambda^R_{pp\mathcal{M}}  \mathfrak{G}_\mathcal{M}(U,V;\sigma,\tau) \\
&\quad+\sum_{ t,\ell,I} \left[\lambda^{R}_{2,t,\ell,I} \lambda^{R}_{p,t,\ell,I}+\lambda^{(0)}_{2,t,\ell,I} \lambda^{(0)}_{p,t,\ell,I}\gamma_{t,\ell,I}^R(\partial_t^\text{no-log}+\frac12\log U)\right]  \mathfrak{G}_{t+\ell,\ell}(U,V;\sigma,\tau) \,.
}
Here, the first line includes the protected multiplets, and the OPE coefficient for the stress tensor multiplet was written explicitly using \eqref{cTolam}. In the second line we denote the singlet long multiplet superblock by $\mathfrak{G}_{\Delta,\ell}$ and $\partial_t^\text{no-log}  \mathfrak{G}_{t+\ell,\ell}(U,V;\sigma,\tau) $ denotes that we consider the term after taking the derivative that does not include a $\log U$, which has already been written separately. The expansion of $\cG_{p}^{R^4}$ takes the same form with $R\to R^4$, except the stress tensor block does not appear.

While the superblock expansion is best expressed in position space, in the large $c_T$ expansion it is also useful to consider the Mellin transform $M_p(s,t;\sigma,\tau)$ of the connected correlator $\mathcal{G}^\text{con}_{p}(U,V;\sigma,\tau)\equiv\mathcal{G}_{p}(U,V;\sigma,\tau)-\mathcal{G}^{(0)}_{p}(U,V;\sigma,\tau)$, which is defined as \cite{Zhou:2017zaw}:
\es{mellinH}{
\mathcal{G}^\text{con}_{p}(U,V;\sigma,\tau)&=\int\frac{ds\, dt}{(4\pi i)^2} U^{\frac s2}V^{\frac t2-\frac{p}{4}-\frac12}M_p(s,t;\sigma,\tau) \\
&\qquad\qquad\times\Gamma\left[\frac p2-\frac s2\right]\Gamma\left[1-\frac s2\right]\Gamma^2\left[\frac p4+\frac12-\frac t2\right]\Gamma^2\left[\frac p4+\frac12-\frac {{u}}{2}\right],\\
}
where $u = p+2 - s - t$ and the integration contours here is defined to include all poles of the Gamma functions on one side of the contour. The Mellin amplitude is defined such that a bulk contact Witten diagram coming from a vertex with $2m$ derivatives gives rise to a polynomial in $s,t$ of degree $m$, and similarly an exchange Witten diagrams corresponds to a Mellin amplitude with poles for the twists of each exchanged operator. The Mellin amplitude must also obey the crossing relations
 \es{crossM}{
 M_p(s,t;\sigma,\tau)  =  M_p(s,u;\tau,  \sigma)   \,, \qquad   M_2(s,t;\sigma,\tau) =    \tau^2M_2(t,s;\sigma/\tau,1/\tau)\,,
 }
which follow from interchanging the first and second operators, and, for $p = 2$, the first and third.  Lastly, $M_p(s,t;\sigma,\tau)$ must satisfy the Ward identities \eqref{ward}, which can be implemented in Mellin space as shown in \cite{Zhou:2017zaw}. Using all these constraints, $M_p(s, t)$ can be expanded similar to the position space expression \eqref{Hlarge} to get
\es{Mplarge}{
M_{p}(s,t)=c_T^{-1}M^R_{p}+c_T^{-\frac53}B^{R^4}(p)M^{R^4}_{p}+\dots\,,
}
where $M^{R^4}_{p}$ is a complicated degree 4 polynomial in $s,t$ whose explicit form we give in the attached \texttt{Mathematica} notebook, while the tree level supergravity amplitude $M^R_{p}$ was written in \cite{Alday:2020dtb} as an infinite sum of the supergravity multiplet and its descendents:
\es{Ms}{
M^R_p=& \sum_{m=0}^\infty\Bigg[  \frac{2^{2 m+5} p ((p-2 t+2) (p+2 (s+2) \sigma -2 (s+t-1))-2 (s+2) \tau  (p-2 (s+t-1)))}{\pi ^2 (2 m+3) (2 m-s+1) \Gamma
   \left(\frac{1}{2}-m\right) \Gamma (2 m+2) \Gamma \left(\frac{1}{2} (-2 m+p-1)\right)} \\
   &+\frac{16 p \tau  \Gamma \left(\frac{p}{2}+1\right) ((p+2 t+2) (p (2 \sigma -1)+2 (-\sigma  s+s+t-1))+2 \tau  (p-s) (p-2
   (s+t-1)))}{\pi  \Gamma \left(\frac{1}{2}-m\right)^2 \Gamma (m+1) \Gamma \left(\frac{p-1}{2}\right) (4 m+p-2 t) \Gamma
   \left(m+\frac{p+3}{2}\right)}\\
   &+\frac{2^{2 m+5} p ((p-2 t+2) (p+2 (s+2) \sigma -2 (s+t-1))-2 (s+2) \tau  (p-2 (s+t-1)))}{\pi ^2 (2 m+3) (2 m-s+1) \Gamma
   \left(\frac{1}{2}-m\right) \Gamma (2 m+2) \Gamma \left(\frac{1}{2} (-2 m+p-1)\right)}\Bigg]\,,
}
where the overall coefficient was fixed by extracting the stress tensor OPE coefficient and comparing to \eqref{cTolam}. Note that this expression is independent of $k$. For even $p$, the sum can be performed to get an expression in terms of Gamma functions, which can then be written in terms of a finite sum of $\bar{D}_{r_1,r_2,r_3,r_4}(U,V)$ functions using its Mellin space definition in Appendix \ref{block3dApp}, as was pointed out in a similar case in \cite{Binder:2021cif}. For instance, for $p=2$ first we resum to get
\es{M2R}{
M^R_2=&\frac{32 \tau  \left(\frac{4 \Gamma \left(\frac{1}{2}-\frac{t}{2}\right)}{\Gamma \left(1-\frac{t}{2}\right)}-\sqrt{\pi } (t+4)\right) ((t+2) (-\sigma  s+s+2 \sigma +t-2)+(s-2) \tau  (s+t-2))}{\pi
   ^{5/2} t (t+2)}\\
&+\frac{32 \sigma  \left(\sqrt{\pi } (s+t-8)+\frac{4 \Gamma \left(\frac{1}{2} (s+t-3)\right)}{\Gamma \left(\frac{1}{2} (s+t-2)\right)}\right) ((t-2) (-\sigma  s+s+2 \sigma +t-6)+(s-2) \tau 
   (s+t-6))}{\pi ^{5/2} (s+t-6) (s+t-4)}\\
&   +\frac{32 \left(\frac{4 \Gamma \left(\frac{1}{2}-\frac{s}{2}\right)}{\Gamma \left(1-\frac{s}{2}\right)}-\sqrt{\pi } (s+4)\right) ((t-2) (-(s+2) \sigma +s+t-2)+(s+2) \tau  (s+t-2))}{\pi ^{5/2} s (s+2)}
\,.
}
Note that the Gamma functions that appear in the denominator exactly cancel one of the Gamma functions in the Mellin transform \eqref{mellinH}, while the poles just shift the arguments of the Gamma functions. We can thus use \eqref{dbarM} to write this expression in terms of a finite number of $\bar{D}(U,V)$ functions. For instance, the contribution to the $[0040]$ irrep in \eqref{Ybasis} is
\es{p2A22R}{
A^R_{[0040]}(U,V)&=-\frac{32 U^3 }{3 \pi ^{5/2} V^2} \Big[ 2 \sqrt{\pi } V^2
   \bar{D}_{1,3,-1,1}(U,V)-2 V^2 \bar{D}_{1,\frac{5}{2},-1,\frac{1}{2}}(U,V)\\
   &+\sqrt{\pi } V^2 \bar{D}_{1,3,0,2}(U,V)
   +\sqrt{\pi } V^2
   \bar{D}_{2,3,-1,2}(U,V)-2 V^{\frac32} \bar{D}_{1,\frac{5}{2},\frac{1}{2},-1}(U,V)\\
   &+3
   \sqrt{\pi } \bar{D}_{3,1,-1,1}(U,V)-\sqrt{\pi } \bar{D}_{4,1,-1,2}(U,V)\Big]\,.
}
While the appearance of half integer and negative arguments might seem nonstandard, in fact these have standard expansions to all orders in $U$ and $V$, including a $\log U$ term, as shown in \cite{Dolan:2000ut} and reviewed in Appendix \ref{block3dApp}. We can then easily expand this and the other channels in superblocks to get the average anomalous dimensions $\langle \lambda^{(0)}_{2,t,\ell} \lambda^{(0)}_{p,t,\ell}\gamma_{t,\ell}\rangle\equiv\sum_{ I} \lambda^{(0)}_{2,t,\ell,I} \lambda^{(0)}_{p,t,\ell,I}\gamma_{t,\ell,I}$ weighted by OPE coefficients for $p=2$:
\es{averageSG}{
\langle \lambda^{(0)}_{2,t,\ell} \lambda^{(0)}_{2,t,\ell}\gamma^R_{t,\ell}\rangle=&-\frac{64 }{\pi ^2} \frac{\sqrt{\pi } (2 \ell+1) \Gamma \left(\frac{t-1}{2}\right) \Gamma \left(\ell+\frac{t}{2}\right) \Gamma (\ell+t+3)}{8 \Gamma \left(\frac{t}{2}\right) \Gamma
   \left(\ell+\frac{t}{2}+\frac{1}{2}\right) \Gamma \left(\ell+t+\frac{5}{2}\right)}\\
&\times\left(\left(2 \ell^2+2 \ell t+6 \ell+t^2+5 t+4\right) \left(\psi({\ell+t+3})-\psi({\ell+1})\right)+(-t-2) (2 \ell+t+3)\right)\,,
}
which matches the values originally computed in \cite{Chester:2018lbz}. We can similarly compute $\langle \lambda^{(0)}_{2,t,\ell} \lambda^{(0)}_{p,t,\ell}\gamma^R_{t,\ell}\rangle$ for higher even $p$, and we show the results for  $p\leq36$ in the attached \texttt{Mathematica} notebook. Since the $M_p^R$ did not depend on $k$, these average anomalous dimensions are also the same for $k=1,2$.

For the $R^4$ amplitude, we need to fix the overall coefficient $B^{R^4}(p)$, which will depend nontrivially on $k$ unlike $M_p^R$. For $p=2$ this was fixed using localization in \cite{Chester:2018aca}, and this derivation could in principle be extended to $p>2$ for $k=1$ using the localization results from \cite{Gaiotto:2020vqj}. For $k=2$ and $p>2$ however, which is our primary interest, we can only fix the coefficient by comparing to the known 11d M-theory S-matrix term in the flat space limit, as was done for the 6d $(2,0)$ theory in \cite{Alday:2020tgi}. In particular, the 11d M-theory S-matrix $\cA$ can be expanded at small Planck length $\ell_{11}$ as
\es{A}{
\mathcal{A}(s,t)=\ell_{11}^{9}\mathcal{A}_{R}+\ell^{15}_{11}\mathcal{A}_{R^4}+\ell^{18}_{11}\mathcal{A}_{R|R}+\ell^{21}_{11}\mathcal{A}_{D^6R^4}+\ell^{23}_{11}\mathcal{A}_{D^8R^4}+\ell_{11}^{24}\mathcal{A}_{R|R^4}+\dots\,,
}
where $s,t,u$ are 11d Mandelstam variables. The lowest few terms $\cA_R$, $\cA_{R^4}$, and $\cA_{D^6R^4}$ are protected, and so can be computed from Type IIA string theory by compactifying on a circle \cite{Green:1997as, Russo:1997mk, Green:2005ba}\footnote{$\cA_{D^4R^4}$ can also be computed in this way, but it vanishes and so we did not write it. Also, the 1-loop supergravity term $\cA_{R|R}$ was computed in \cite{Russo:1997mk,Green:1997as}, while $\cA_{R|R^4}$ and $\cA_{R^4|R^4}$ were computed in \cite{Alday:2020tgi}.} to get
  \es{SGtoR4}{
  \frac{ \mathcal{A}_{R^4}}{\mathcal{A}_{R}}=\frac{stu}{3\cdot 2^7}\,,\qquad  \frac{ \mathcal{A}_{D^6R^4}}{\mathcal{A}_{R}}=\frac{(stu)^2}{15\cdot  2^{15}}\,.
   }
The small $\ell_{11}$ expansion in 11d maps to the large $N$ expansion in the CFT according to the dictionary
   \cite{Aharony:2008ug,Aharony:2008gk}:
\es{cPlanck}{
  \frac{L^6}{\ell_{11}^6}=\left(\frac{3\pi c_T k}{2^{11}}\right)^{\frac23}+O(c_T^0) \,.
}
The flat space limit formula \cite{Penedones:2010ue,Chester:2018aca} then relates a Mellin amplitude $M^a_p(s,t)$ of large $s,t$ degree $a$ to the 11d amplitude defined in \eqref{A} as
 \es{flat}{
c_T^{\frac{2(1-a)}{9}}\frac{\pi ^{5/2} 2^{-a-p-3} \Gamma (p-1)}{\Gamma \left(\frac{1}{2} (2 a+p-1)\right)}\lim_{s,t\to\infty} \frac{s t (s+t)M^a(s,t)}{(t (-\sigma  s+s+t)+s \tau  (s+t))^2}= \ell_{11}^{2a-2}\frac{{\mathcal{A}_{2a+7}}}{\cA_R}\,,
 }
 where $\mathcal{A}_{2a+7}$ is a term in the amplitude with length dimension $(2a+7)$, and $\ell_{11}$ is the 11d Planck length. For instance, $\mathcal{A}_{15}\equiv\mathcal{A}_{R^4}$ has length dimension 15 in \eqref{A} and corresponds to $M_p^4\equiv M_p^{R^4}$. The 11d amplitude of course is the same for all $p$, and the ratio of $\mathcal{A}_{R^4}/\mathcal{A}_R$ was given in \eqref{SGtoR4}. Using this 11d amplitude and the flat space limit we find 
 \es{BR4}{
B^{R^4}(p)=\frac{32\cdot{2^{\frac13}} (p-1) (p+1) (p+3) (p+5)}{3\ 3^{2/3} \pi ^{8/3} k^{2/3} \Gamma \left(\frac{p}{2}\right)}\,,
}
 which matches the $p=2$ result computed from localization in \cite{Chester:2018aca}. Note that the simple $k$ dependence comes from the AdS/CFT dictionary \eqref{cPlanck}. Since $M_p^{R^4}$ is just a polynomial Mellin amplitude for every $p$, we can convert to a finite number of $\bar{D}(U,V)$ to get
 \es{p2A22R4}{
A^{R^4}_{[0040]}(U,V)&=\
 \frac{64 \cdot{2^{\frac13}} \left(p^2-1\right) U^{p/2}} { 3^{8/3} \pi ^{8/3} k^{2/3} \Gamma \left(\frac{p}{2}\right)}
  \Big[6 (p-2) p \bar{D}_{\frac{p}{2},\frac{p}{2},1,1}(U,V)+4 (p-3) (p-2) (p+2)
   \bar{D}_{\frac{p}{2},\frac{p}{2},2,2}(U,V)\\
   &+192 \bar{D}_{\frac{p}{2},\frac{p}{2},3,3}(U,V)-312
   \bar{D}_{\frac{p}{2},\frac{p}{2},4,4}(U,V)+ 
  p (p-1) ((p-1) p-26)
   \bar{D}_{\frac{p}{2},\frac{p}{2},3,3}(U,V)\\
   &+4 p(p (p+2)-29) \bar{D}_{\frac{p}{2},\frac{p}{2},4,4}(U,V)+4p (p+8)
   \bar{D}_{\frac{p}{2},\frac{p}{2},5,5}(U,V)+60 \bar{D}_{\frac{p}{2},\frac{p}{2},5,5}(U,V)
   \Big]\,,
 }
 and similarly for the other channels. We can then expand in superblocks to get the even $p$ average anomalous dimensions 
\es{averageR4}{
\langle \lambda^{(0)}_{2,t,\ell} \lambda^{(0)}_{p,t,\ell}\gamma^{R^4}_{t,\ell}\rangle=&
\delta_{\ell,0}(-1)^{\frac p2} \frac{ (t+1) 2^{p+t+\frac{25}{3}} \Gamma \left(\frac{t}{2}+2\right) \Gamma \left(\frac{t}{2}+4\right) \Gamma \left(\frac{1}{2} (p+t+5)\right)}{ 3^{5/3} \pi ^{8/3}
   k^{2/3} \Gamma (p-1) \Gamma \left(t+\frac{5}{2}\right) \Gamma \left(\frac{1}{2} (-p+t+2)\right)}
\,,
}
which are only nonzero for zero spin, and for $p=2$ match the results in \cite{Chester:2018aca}.

\subsection{Large $c_T$ expansion of $\langle2222\rangle$}
\label{2222largec}

Finally, we restrict to the stress tensor correlator $\langle2222\rangle$, which is our primary interest. For simplicity, we will drop the $p=2$ subscript from all further expressions. The Mellin amplitude $M(s,t)$ is fixed by the analytic structure, growth at infinity, and crossing symmetry to take the form \eqref{M2222} given in the Introduction, where the coefficient of each $c_T^{-b}$ must include all allowed Mellin amplitudes of large $s,t$ degree $(9/2b-7/2)$ or less. These can include the polynomial Mellin amplitudes $M^a$ given in \cite{Chester:2018aca}, which we also include in the attached \texttt{Mathematica} file. For the amplitudes we consider above, this implies one allowed polynomial Mellin amplitude of each degree $a$, which correspond to contact Witten diagrams with $2a$ derivatives. These contact diagrams only contribute to a finite number of spins that grows with the degree \cite{Heemskerk:2009pn}. For the multiplets in $\langle2222\rangle$, the contribution from each $M^a$ in \eqref{M2222} is summarized in Table 4 of \cite{Chester:2018aca}, which we repeat here in Table \ref{resultList}. The other Mellin amplitudes shown in \eqref{M2222} include the tree level supergravity term $M^R$ discussed in the previous section, which includes poles for the single trace supergravity multiplet and its descendents, as well as the 1-loop Mellin amplitudes $M^{R|R}$ and $M^{{R^4}|R}$ of degrees $5.5$ and $8.5$, respectively, that we will discuss more in the following section. 

\begin{table}
\centering
\begin {tabular} {| c || c | c | c | c | }
\hline
 {CFT data:}&$M^{4}$  &$M^{6}$ & $M^{7}$ &$M^{8}$  \\
  \hline
\TBstrut  $\lambda^2_{(B,+)}$& $ \frac{256}{35} $ & $-\frac{59392}{693}$ & $-\frac{477184}{429}$ & $\frac{4022091776}{4448925}$ \\
  \hline
  \TBstrut  $\lambda^2_{(B,2)}$& $  \frac{256}{7} $& $-\frac{296960}{693}$ & $-\frac{2385920}{429}$ & $\frac{4022091776}{889785}$ \\
  \hline
  \TBstrut  $\lambda^2_{(A,+)_0}$&  0& $\frac{16384}{1485}$ & $\frac{950272}{6435}$ & $-\frac{131396796416}{467137125}$ \\
  \hline
 \TBstrut  $\lambda^2_{(A,+)_2}$& 0& $0$ & $0$ & $\frac{67108864}{557375}$ \\
  \hline
\TBstrut  $\lambda^2_{(A,2)_1}$& 0& $\frac{131072}{1155}$ & $\frac{21889024}{15015}$ & $-\frac{3848847491072}{1089986625}$ \\
  \hline
 \TBstrut  $\lambda^2_{(A,2)_3}$& 0& $0$ & $0$ & $\frac{268435456}{121275}$ \\
  \hline
 \TBstrut $\gamma_{2,0}$& -192& $\frac{15360}{11}$ & $\frac{192000}{11}$ & $-\frac{18059264}{1521}$ \\
  \hline
 \TBstrut   $\gamma_{2,2}$& 0& $-1536$ & $-18432$ & $\frac{509591552}{12675}$ \\
  \hline
 \TBstrut     $\gamma_{2,4}$& 0& $0$& $0$ & $-32768$ \\
  \hline
\end{tabular}
\caption{Contributions from large $s,t$ degree $a$ contact Mellin amplitudes $M^{a}(s,t)$ to the OPE coefficients squared $\lambda^2_{22\cM}$ of some protected multiplets $(B,+)$, $(B,2)$, $(A,+)_\ell$ for even $\ell$, and $(A,2)_\ell$ for odd $\ell$, as well as to the anomalous dimensions $\g_{t,\ell}$ for even $\ell$ of the lowest twist $t=2$ unprotected multiplet $(A,0)_{t+\ell,\ell}$. Adapted from \cite{Chester:2018aca} with typos fixed.
}\label{resultList}
\end{table}

The coefficient $B^{R^4}_4\equiv B^{R^4}(2)$ was fixed in the previous section. The three $B^{D^6R^4}$ coefficients were fixed in  \cite{Binder:2018yvd} from the two localization constraints and the flat space limit \eqref{flat} to get
\es{D6R4answer}{
B^{D^6R^4}_4=-\frac{1352960\ 6^{2/3}}{9 \pi ^{10/3}k^{\frac43}}\,,\qquad B^{D^6R^4}_6=-\frac{220528\ 6^{2/3}}{\pi ^{10/3}k^{\frac43}}\,,\qquad B^{D^6R^4}_7=\frac{16016\ 6^{2/3}}{\pi ^{10/3}k^{\frac43}}\,.
} 
There are not enough constraints to fully fix the other tree amplitudes, while the 1-loop amplitudes will be considered in the next section, so for now we will only extract CFT data up to order $c_T^{-\frac73}$ and leave the $R|R$ term unknown for now. The short multiplets OPE coefficients $\lambda^2_{(B,+)}$ and $\lambda^2_{(B,+)}$ were in fact computed to all orders in $1/c_T$ using localization in \cite{Agmon:2017xes}, and take the form for $k=2$:
\es{B2Bp}{
\lambda^2_{(B,2)}&=\frac{32}{3}-\frac{1024 \left(4 \pi ^2-15\right)}{9 \pi ^2
   c_T}+\frac{20480 \left(\frac{2}{3}\right)^{\frac23}}{\pi ^{8/3} c_T^{5/3}}+\frac{16384
   \left(2 \pi ^2-25\right)}{9 \pi ^4 c_T^2}-\frac{327680 ({\frac{2}{3}})^{\frac13}}{3
   \pi ^{\frac{10}{3}} c_T^{7/3}}+\frac{7536640 \left(\frac{2}{3}\right)^{\frac23}}{9 \pi
   ^{\frac{14}{3}} c_T^{8/3}}+O(c_T^{-3})\,,\\
   \lambda^2_{(B,+)}&=\frac{16}{3}+\frac{1024 \left( \pi ^2+3\right)}{9 \pi ^2
   c_T}+\frac{4096 \left(\frac{2}{3}\right)^{2/3}}{\pi ^{8/3} c_T^{5/3}}+\frac{16384
   \left(2 \pi ^2-25\right)}{45 \pi ^4 c_T^2}-\frac{65536
(\frac{2}{3})^{\frac13}}{3 \pi ^{\frac{10}{3}} c_T^{7/3}}+\frac{1507328
   \left(\frac{2}{3}\right)^{2/3}}{9 \pi ^{14/3} c_T^{8/3}}+O(c_T^{-3})\,,\\
}
where note that these OPE coefficients are related due to crossing in the 1d topological sector as \cite{Chester:2014mea}
\es{1drel}{
\frac{1024}{c_T}-5\lambda^2_{(B,+)}+\lambda^2_{(B,2)}+16=0\,,
}
so in fact only one is independent. For the other multiplets in $S_2\times S_2$ we get for $k=1,2$
\es{2222data}{
\lambda^2_{(A,+)_\ell}&=\frac{\pi  \Gamma (\ell+3)^2}{\Gamma \left(\ell+\frac{5}{2}\right)^2}-\frac{64 \Gamma (\ell+3)^2 (-2 \ell+2 (\ell (\ell+5)+5) \psi ^{(1)}(\ell+3)-5)}{\pi c_T\Gamma \left(\ell+\frac{5}{2}\right)^2}\\
&+c_T^{-2}(\lambda^{R|R}_{(A,+)_\ell})^2-\delta_{\ell,0}\frac{1835008\ 6^{2/3}}{27 \pi ^{10/3}c_T^{7/3} k^{4/3}}+O(c_T^{-\frac{23}{9}})\,,\\
\lambda^2_{(A,2)_\ell}&=\frac{\pi  \Gamma (\ell+2) \Gamma (\ell+4)}{\Gamma \left(\ell+\frac{3}{2}\right) \Gamma \left(\ell+\frac{7}{2}\right)}+\frac{64 \Gamma (\ell+4)^2 \left((2 \ell+3) (\ell (\ell+7)+11)-2 (\ell+2)^2 (\ell+3)^2 \psi ^{(1)}(\ell+2)\right)}{\pi c_T(\ell+2)^2 (\ell+3)^2 \Gamma \left(\ell+\frac{3}{2}\right) \Gamma
   \left(\ell+\frac{7}{2}\right)}\\
   &+c_T^{-2}(\lambda^{R|R}_{(A,2)_\ell})^2-\delta_{\ell,1}\frac{8388608\ 6^{2/3}}{5 \pi ^{10/3}c_T^{7/3} k^{4/3}}+O(c_T^{-\frac{23}{9}})\,,\\
\Delta_{2,\ell}&=2+\ell-\frac{256 (2 \ell+3) (2 \ell+5) (2 \ell+7)}{\pi ^2c_T (\ell+1) (\ell+2) (\ell+3) (\ell+4)}-\frac{71680\cdot{6}^{\frac13} \delta _{0,\ell}}{\pi ^{8/3}c_T^{5/3} k^{2/3}}\\
&+c_T^{-2}\gamma_{2,\ell}^{R|R}+c_T^{-\frac73}\left(\delta_{\ell,0}\frac{1433600\ 6^{2/3}}{3 \pi ^{10/3}  k^{4/3}}+\delta_{\ell,2}\frac{43524096\ 6^{2/3}}{\pi ^{10/3} k^{4/3}}\right)+O(c_T^{-\frac{23}{9}})\,,\\
}
where $\ell$ is even for $\Delta_{2,\ell}$ and $\lambda^2_{(A,+)_\ell}$, odd for $\lambda^2_{(A,2)_\ell}$, and for $\Delta_{2,\ell}$ we only wrote the result for the lowest twist because recall that higher twists are degenerate and so require unmixing beyond leading order. In the following section, we will determine the 1-loop corrections to some of this non-trivial CFT data.

\section{$\langle2222\rangle$ at 1-loop}
\label{1loop}

We now discuss the 1-loop terms $R|R$ at $c_T^{-2}$, $R|R^4$ at $c_T^{-\frac83}$, and $R^4|R^4$ at $c_T^{-\frac{10}{3}}$ for $k=2$ ABJ(M) theory. For each term we compute the double-discontinuity (DD) from the tree and GFFT data derived in the previous sections, and then use it as well as crossing symmetry and the superconformal Ward identity to write the entire correlator in Mellin space  up to contact term ambiguities. We then take the flat space limit and match these correlators to the relevant 1-loop corrections to the 11d S-matrix. Finally, we extract low-lying CFT data using two methods: the Lorentzian inversion integral applied to the DD \cite{Alday:2016njk,Caron-Huot:2017vep} and a projection method applied to the entire Mellin amplitude \cite{Heemskerk:2009pn,Chester:2018lbz}. The inversion method does not converge for low spins that are affected by contact term ambiguities, while the projection method can be used to compute that CFT data in terms of those ambiguities. In the next section, we will discuss how to use localization and a conjectured analytic continuation of the inversion method to fix all the contact term ambiguities for $R|R$ and $R|R^4$.

\subsection{One-loop from tree level}
\label{1loopfrom}

We begin by expanding the correlator $\cG$ for $\langle2222\rangle$ to 1-loop order at large $c_T$ using the block expansion described in section \ref{qqpp}. For $R|R$ at order $c_T^{-2}$, this takes the form
\es{RR}{
\cG^{R|R}=&\sum_{t=2,4,\dots}\sum_{\ell\in\text{Even}}\Big[\frac18\langle(\lambda^{(0)}_{t,\ell})^2(\gamma^R_{t,\ell})^2\rangle(\log^2U+4\log U\partial_t^\text{no-log}+4(\partial_t^\text{no-log})^2) \\
&+\frac12\langle(\lambda^{R})^2_{t,\ell}\gamma^{R}_{t,\ell}\rangle(\log U+2\partial_t^\text{no-log})\\
 &+\frac12\langle(\lambda^{(0)}_{t,\ell})^2\gamma^{{R}|{R}}_{t,\ell}\rangle(\log U+2\partial_t^\text{no-log})+\langle(\lambda^{{R}|{R}}_{t,\ell})^2\rangle\Big] \mathfrak{G}_{t+\ell,\ell}(U,V;\sigma,\tau)\\
 &+\sum_{\mathcal{M}_{\Delta,\ell}\in\{(B,+),(B,2),(A,2)_\ell,(A,+)_\ell\}}(\lambda^{R|R}_{22\mathcal{M}} )^2   \mathfrak{G}_\mathcal{M}(U,V;\sigma,\tau)\,,
}
where $\partial_t^\text{no-log} \mathfrak{G}_{t+\ell,\ell}(U,V;\sigma,\tau)$ was defined in \eqref{SGexp}. The first three lines describe the double trace singlet long multiplets $(A,0)^{[0000]}_{t+\ell,\ell}$, where $\langle\rangle$ denotes the average over the $ (t-1)$-fold degenerate operators. The fourth line includes all the protected multiplets in $\langle 2222\rangle$ except the stress tensor multiplet, which is $1/c_T$ exact. The expression for $\cG^{R^4|R^4}$ at order $c_T^{-\frac{10}{3}}$ is identical except we replace $R\to R^4$ and the sum for the long multiplets is now restricted to $\ell=0$, while for $\cG^{R|R^4}$ at order $c_T^{-\frac83}$ we furthermore replace the $\frac18$ in the first line by $\frac14$, since the vertices are different.

As shown in \cite{Aharony:2016dwx}, the entire 1-loop term up to the contact term ambiguities described in Section \ref{2222largec} can in fact be constructed from the $\log^2 U$ terms shown above, which are written in terms of GFFT and tree data, since under $1\leftrightarrow3$ crossing 
\es{crossing}{
\cG(U,V;\sigma,\tau)=\frac{U}{V}\tau^2 \cG(V,U;\sigma/\tau,1/\tau)\,,
}
 the $\log^2U$ terms are related to $\log^2V$ terms that are the only contributions at this order to the DD, which fixes the entire correlator according to the Lorentzian inversion formula \cite{Alday:2016njk,Caron-Huot:2017vep}. Note that the average $\langle(\lambda^{(0)}_{t,\ell})^2\gamma^A_{t,\ell}\gamma^{B}_{t,\ell}\rangle$ for 1-loop vertices $A,B$ is what appears in the $\log^2U$ term, whereas the different averages $\langle(\lambda^{(0)}_{t,\ell})^2\gamma^A_{t,\ell}\rangle$ and $\langle(\lambda^{(0)}_{t,\ell})^2\gamma^B_{t,\ell}\rangle$ are what appear at tree level. As shown in \cite{Alday:2017xua,Aprile:2017bgs,Aprile:2017xsp,Alday:2018pdi} for $\mathcal{N}=4$ SYM and \cite{Alday:2020tgi} for 6d $(2,0)$, one can compute $\langle(\lambda^{(0)}_{t,\ell})^2\gamma^A_{t,\ell}\gamma^{B}_{t,\ell}\rangle$ from GFFT $\langle ppqq\rangle$ and tree level $\langle22pp\rangle$ data as
\es{appA}{
\langle(\lambda^{(0)}_{t,\ell})^2\gamma^A_{t,\ell}\gamma^{B}_{t,\ell}\rangle=\sum_{p=2,4,\dots}^{t}\frac{\langle\lambda^{(0)}_{2,t,\ell}\lambda^{(0)}_{p,t,\ell}\gamma^A_{t,\ell}\rangle  \langle\lambda^{(0)}_{2,t,\ell}\lambda^{(0)}_{p,t,\ell}\gamma^B_{t,\ell}\rangle}{ {\langle(\lambda^{(0)}_{p,t,\ell})^2\rangle} }\,,
}
where we summed over each $p$ for which a given twist $t$ long multiplet appears. Unlike the 4d and 6d cases, in 3d the sum only runs over even $p$ regardless of the orbifold.\footnote{As discussed in the introduction, for $k=1$ ABJM the DD would receive additional contributions from the OPE coefficients of odd twist long multiplets that appear for odd $p$.} We computed ${\langle(\lambda^{(0)}_{p,t,\ell})^2\rangle} $, $\langle\lambda^{(0)}_{2,t,\ell}\lambda^{(0)}_{p,t,\ell}\gamma^R_{t,\ell}\rangle$, and $\langle\lambda^{(0)}_{2,t,\ell}\lambda^{(0)}_{p,t,\ell}\gamma^{R^4}_{t,\ell}\rangle$ in \eqref{ppppLam}, \eqref{averageSG}, and \eqref{averageR4}, respectively, which is sufficient to compute $R|R$, $R|R^4$, and $R^4|R^4$ for the $k=2$ ABJ(M) theory. The $p,t,\ell$ sums for the $\log^2U$ term in $R|R$ can be done by expanding at small $U$ in each $R$-symmetry channel to get:
\es{slices}{
\frac18\sum_{t=2,4,\dots}& \sum_{\ell\in\text{Even}}\sum_{p=2,4,\dots}^{t} \frac{\langle\lambda^{(0)}_{2,t,\ell} \lambda^{(0)}_{p,t,\ell}\gamma^R_{t,\ell}\rangle^2  }{ {\langle(\lambda^{(0)}_{p,t,\ell})^2\rangle} } \mathfrak{G}_{t+\ell,\ell}(U,V;\sigma,\tau)=\\
& Y_{[0000]}(\sigma,\tau) \big[U h_{R|R}^{(1),[0000]}(V) + \cdots \big]+Y_{[0100]}(\sigma,\tau) \big[U h_{R|R}^{(1),[0100]}(V) + \cdots \big]\\
& Y_{[0020]}(\sigma,\tau) \big[U h_{R|R}^{(1),[0020]}(V) + \cdots \big]+Y_{[0120]}(\sigma,\tau) \big[U^2 h_{R|R}^{(2),[0120]}(V) + \cdots \big]\\
& Y_{[0200]}(\sigma,\tau) \big[U^2 h_{R|R}^{(2),[0200]}(V) + \cdots \big]+Y_{[0040]}(\sigma,\tau) \big[U^3 h_{R|R}^{(3),[0040]}(V) + \cdots \big]\,,
}
where the different powers of $U$ in each channel correspond to the lowest twists that appear in each channel for the $(A,0)_{t+\ell,\ell}^{[0000]}$ superblock as given in Table 8 of \cite{Chester:2014fya}. The $U$-slices in each channel take the form
\es{slices2}{
h^{(n),[0ab0]}_{R|R}(V) ={P^{[0ab0]}_{1,R|R}(V)}\log ^2 V + {P^{[0ab0]}_{2,R|R}(V)} \text{Li}_2(1-V)+ {P^{[0ab0]}_{3,R|R}(V)} \log V  + {P^{[0ab0]}_{4,R|R}(V)} \,,
}
where $P^{[0ab0]}_{i,R|R}(V)$ are polynomials in $V$ divided by monomials of $(1-V)$ whose precise degree varies in each channel. The expressions for $R|R^4$ and $R^4|R^4$ are also given by \eqref{appA}, except the $\ell$ sum is trivially $\ell=0$ in those cases, and $R|R^4$ has an extra factor of 2. The $U$-slices for $R|R^4$ take the simpler form
\es{slices3}{
h^{(n),[0ab0]}_{R|R^4}(V) = {P^{[0ab0]}_{1,R|R^4}(V)}\log V  + {P^{[0ab0]}_{2,R|R^4}(V)}\,,
}
for similarly defined polynomials divided by monomials, and a similar expression holds for $R^4|R^4$. We give the explicit expressions for many $n$ in the attached \texttt{Mathematica} file. 

\subsection{Mellin amplitude and comparison to 11d}
\label{reducedMellinAmplitude}

We now show how to complete the position space DD to the entire correlator using crossing symmetry and the superconformal Ward identity in Mellin space. For $R|R^4$ and $R^4|R^4$ we will find closed form expressions up to the expected contact term ambiguities, while for $R|R$ we are able to compute a closed form up to a certain polynomial in $s,t$ that in principle can be fixed from the Ward identity, but is difficult to fix in practice. The expressions we find in all cases are sufficient to check the flat space limit comparison to 11d, as well as to extract all CFT data except for some low spins.

We can compute the Mellin amplitudes from the resummed DD's following a similar but more complicated version of the calculation in the 4d \cite{Alday:2018kkw} and 6d \cite{Alday:2020tgi} cases. In the previous section, we computed the coefficient of $\log^2U$ in the $s$-channel, which gave the DD in the $t$-channel as an expansion in small $U$. From the definition of the Mellin transform in \eqref{mellinH}, we can then convert $U^n\log^2U h^{(n)}(V)$ to an $s$-pole in $M(s,t)$ as
\es{res}{
 U^{n}  \log^2 U  h_n(V) \leftrightarrow \frac{res_{n-1}(t)}{s-2(n-1)}\,,
}
where the residues $res_{n-1}(t)$ follows from the $t$-integral in \eqref{mellinH}. For $R^4|R$ and $R^4|R^4$, $res_{n-1}(t)$ is analytic because the slices \eqref{slices3} do not contain any $\log^2V$ terms. In these cases, we can then use crossing symmetry \eqref{crossM} to fix the other parts of the Mellin amplitude that are analytic in $s$ to get
\es{RR4c}{
M^{R|R^4}(s,t;\sigma,\tau) &= \sum_{m=1}^\infty\Bigg[ \frac{1}{(s-2m)}\left(\frac{\hat{c}(m,s,t;\sigma,\tau)}{m}+\frac{\hat{d}(m,s,t;\sigma,\tau)\Gamma(m)}{\Gamma(m+\frac12)}\right) \\
&\qquad\qquad\qquad\qquad\qquad\qquad+\text{crossed}    \Bigg]+\sum_{i=1}^{50} \hat k_i{\bf P}^{(8),i}(s,t;\sigma,\tau)\,,\\
M^{R^4|R^4}(s,t;\sigma,\tau) &= \sum_{m=1}^\infty\left[ \frac{\hat{\hat{d}}(m,s,t;\sigma,\tau)\Gamma(m)}{(s-2m)\Gamma(m+\frac12)} +\text{crossed}    \right]+\sum_{i=1}^{84}\hat{\hat k}_i{\bf P}^{(11),i}(s,t;\sigma,\tau)\,,\\
}
where $\hat{c}(m,s,t;\sigma,\tau)$, $\hat {d}(m,s,t;\sigma,\tau)$, and $\hat{\hat {d}}(m,s,t;\sigma,\tau)$ are quadratic in $\sigma,\tau$ and polynomials in $m,s,t$, while ${\bf P}^{(8),i}(s,t;\sigma,\tau)$ and ${\bf P}^{(11),i}(s,t;\sigma,\tau)$ parameterize all crossing symmetric degree 8 and 11 polynomials in $s,t$, respectively, whose coefficients $\hat k_i$ and $\hat{\hat k}_i$ should be fixed by the superconformal Ward identity in terms of the physical contact term ambiguities in \eqref{M2222}. Note that one can swap $s$ for $2m$ in these expressions to get the same residues at the poles, which only changes the $\hat k_i$ and $\hat{\hat k}_i$. The only rule in performing this swap is that the degree of $M^{R|R^4}(s,t) $ and $M^{R^4|R^4}(s,t) $ at large $s,t$ does not exceed $8.5$ and $11.5$, respectively. In practice we can simply set $s=2m$ and similarly for the crossed terms, which allows us to resum to get
\es{RR4c2}{
M^{R|R^4} &=\Bigg[\frac{\psi \left(1-\frac{s}{2}\right)}{s} {\bf p}_1(s,t;\sigma,\tau)+\frac{\, _3F_2\left(1,1,1-\frac{s}{2};\frac{3}{2},2-\frac{s}{2};1\right)}{2-s}  {\bf p}_2(s,t;\sigma,\tau)\\
&\qquad+\text{crossed}\Bigg]+\sum_{i=1}^{50} \hat k_i{\bf P}^{(8),i}(s,t;\sigma,\tau)\,,\\
M^{R^4|R^4}&=\Big[\frac{\, _3F_2\left(1,1,1-\frac{s}{2};\frac{3}{2},2-\frac{s}{2};1\right)}{2-s}  {\bf p}_3(s,t;\sigma,\tau)+\text{crossed}\Bigg]+\sum_{i=1}^{84} \hat{\hat k}_i{\bf P}^{(8),i}(s,t;\sigma,\tau)\,,\\
}
where ${\bf p}_i(s,t;\sigma,\tau)$ are various polynomials in $s,t,\sigma,\tau$ that are given in the attached \texttt{Mathematica} file, along with the explicit $\hat k_i$ and $\hat{\hat k}_i$ that we fix using the Mellin space Ward identity. At large $s,t$ these amplitudes take the form
\es{flatRR4}{
\lim_{s,t\to\infty}M^{R|R^4} (s,t;\sigma,\tau)&=\frac{655360\cdot {2}^{\frac16} (-s)^{9/2}(t (-\sigma  s+s+t)+s \tau  (s+t))^2}{3^{5/3} \pi ^{19/6}}+\text{crossed}\,,\\
\lim_{s,t\to\infty}M^{R^4|R^4} (s,t;\sigma,\tau)&=\frac{63078400 \cdot  (-s)^{15/2} (t (-\sigma  s+s+t)+s \tau  (s+t))^2}{2^{\frac16}{3}^{\frac43} \pi ^{23/6}}+\text{crossed}\,,
}
where we assumed $s,t<0$. We can then use the flat space limit formula \eqref{flat} for $p=2$ and $a=8.5$ and $a=11.5$ for $M^{R|R^4}$ and $M^{R^4|R^4}$, respectively, as well as the relation between $\ell_{11}$ and $c_T$ in \eqref{cPlanck} for $k=2$, to get precisely the expected 11d amplitudes as given in equations 4.28 and 4.29 of \cite{Alday:2020tgi}.

For $R|R$, the residues $res_{n-1}(t)$ in \eqref{res} now contain poles in $t$ because the slices \eqref{slices2} contain $\log^2V$ terms. The resulting Mellin amplitude thus contains both single and double pole terms
\es{MellinRR}{
M^{R|R}(s,t;\sigma,\tau) &= \Bigg[ \sum_{m,n=1}^\infty \frac{c(m,n,s,t;\sigma,\tau)}{(s-2m)(t-2n)}\frac{\Gamma(m)\Gamma(n)\Gamma(m+n-\frac{11}{2})}{\Gamma(m+n-1)\Gamma(n-\frac12)\Gamma(m-\frac12)} \\
&  + \sum_{m=1} \frac{1}{s-2m}\Bigg(\frac{d(m,s,t;\sigma,\tau)\Gamma(m)}{\Gamma(m+\frac12)(m-4)(m-3)(m-2)(m-1)}\\
&+\frac{e(m,s,t;\sigma,\tau)}{(m-4)(m-3)(m-2)(m-1)(2m-3)(2m-5)(2m-7)(2m-9)}\Bigg)\\
& +\text{crossed}\Bigg]+\hat{\bf P}(s,t;\sigma,\tau)+\sum_{i=1}^{24} k_i{\bf P}^{(5),i}(s,t;\sigma,\tau)\,.
}
Here, ${c}(m,n,s,t;\sigma,\tau)$, $ {d}(m,s,t;\sigma,\tau)$, and ${ {e}}(m,s,t;\sigma,\tau)$ are quadratic in $\sigma,\tau$ and polynomials in $m,s,t$, while ${\bf P}^{(5),i}(s,t;\sigma,\tau)$ are all crossing symmetric degree 5 polynomial in $s,t$, which in principle should be fixed by the superconformal Ward identity in terms of just one of the 24 $k_i$. For the double pole residues we can swap $s$ for $2m$ and $t$ for $2n$ to get the same residue at the poles, but which will change the single pole residues and the $k_i$. When swapping we must be careful that the resulting sums are all finite, and that the large $s,t$ growth does not exceed $5.5$. In fact, for all choices of swaps the large $s,t$ degree exceeds $5.5$, which is why we must also include the polynomial $\hat{\bf P}(s,t;\sigma,\tau)$ that generically will have degree greater than $5.5$, and is fixed to cancel the corresponding large $s,t$ terms from the single and double sum terms. 

It is difficult to check the large $s,t$ limit of \eqref{MellinRR} with the constraints just discussed, because we must compute the double sum term to subleading order in large $s,t$ to take into account the cancellations of the leading terms by $\hat{\bf P}(s,t;\sigma,\tau)$. Instead, we can more easily compute the large $s,t$ limit by considering the ansatz \eqref{MellinRR} but with coefficients with unphysical poles at $s,t,u=0$:
\es{goodflat}{
c_\text{flat}(s,t;\sigma,\tau)&\equiv  \frac{4mn}{st}c(s/2,t/2,s,t;\sigma,\tau)\,,\\
d_\text{flat}(s,t;\sigma,\tau)&\equiv  \frac{1}{t}d(s/2,s,t;\sigma,\tau)\,,\qquad e_\text{flat}(s,t;\sigma,\tau)\equiv  \frac{1}{t}e(s/2,s,t;\sigma,\tau)\,,\\
}
as well as replacing the ${\bf P}^{(5),i}(s,t;\sigma,\tau)$ by a higher degree polynomial multiplied by $\frac{1}{st}$, and without the now unnecessary $\hat{\bf P}(s,t;\sigma,\tau)$. The resulting explicit Mellin amplitude is given in the attached \texttt{Mathematica} file. In principle, we can completely fix the $k_i$ using the superconformal Ward identity, which should cancel all the unphysical poles at $s,t,u=0$. In practice, we checked that all the residues for both the double and single poles in $s,t$ are satisfied by the Ward identity, but did not carefully fix the $k_i$ since the terms they multiply are by definition subleading in the large $s,t$ limit and we only use this formulation to check the flat space limit. In particular, to get the leading large $s,t$ term that appears in the flat space limit formula \eqref{flat}, we should look at the regime where $m,n,s,t$ all scale equally large, in which case we can replace the sums over $m,n$ by integrals. For the double pole term we get
\es{dubPole}{
&\lim_{s,t\to\infty}\sum_{m,n=1}^\infty \frac{c_\text{flat}(m,n,s,t;\sigma,\tau)}{(s-2m)(t-2n)}\frac{\Gamma(m)\Gamma(n)\Gamma(m+n-\frac{11}{2})}{\Gamma(m+n-1)\Gamma(n-\frac12)\Gamma(m-\frac12)}+\text{crossed}\\
&=\int_0^\infty dmdn\frac{5120 m^{3/2} n^{3/2} s t (s+t) (t (-\sigma  s+s+t)+s \tau  (s+t))^2}{3 \pi ^{7/2} (m+n)^{9/2} (s-2 m) (t-2 n)}+\text{crossed}\\
&=-\frac{1024 \sqrt{2} s t(t (-\sigma  s+s+t)+s \tau  (s+t))^2}{63 \pi ^{5/2} (s+t)^4 \sqrt{s t}} \Bigg[6
   (-t)^{9/2}-41 s^2 (-t)^{5/2}+88 s^3 (-t)^{3/2}\\
   &+88 (-s)^{3/2} t^3-8 \sqrt{-s} t^4-45 (-s)^{7/2} t-45 s (-t)^{7/2}+6 (-s)^{9/2}-8 s^4 \sqrt{-t}-41 (-s)^{5/2} t^2\\
 &+105 s^2 t^2 \sqrt{-s-t} \log \Big[\frac{\left(\sqrt{-s-t}+\sqrt{-s}\right) \left(\sqrt{-s-t}+\sqrt{-t}\right)}{\sqrt{s
   t}}\Big]  \Bigg]+\text{crossed}\,,\\
}
where we assumed that $s,t<0$. For the single pole terms the $d_\text{flat}$ term is leading at large $s,t$ and gives
 \es{singPole}{
&\lim_{s,t\to\infty} \sum_{m=1} \frac{1}{s-2m}\frac{d_\text{flat}(m,s,t;\sigma,\tau)\Gamma(m)}{\Gamma(m+\frac12)(m-4)(m-3)(m-2)(m-1)}+\text{crossed}\\
&\qquad=-\int_0^\infty dm \frac{18}{7} \sqrt{\pi } m^{3/2} \left(4 m^2 \tau +2 m t (-\sigma +\tau +1)+t^2\right)^2+\text{crossed}\\
&\qquad=\frac{4096 \sqrt{2}}{7 \pi ^{5/2} }  (t (-\sigma  s+s+t)+s \tau  (s+t))^2 \left(i t \sqrt{-s-t}+i s \sqrt{-s-t}+(-s)^{3/2}+(-t)^{3/2}\right)\,.\\
}
We then plug these large $s,t$ expressions into the flat space limit formula \eqref{flat} for $p=2$ and $a=5.5$ and use the relation between $\ell_{11}$ and $c_T$ in \eqref{cPlanck} for $k=2$ to precisely get the expected 11d amplitudes as given in equations 4.19 of \cite{Alday:2020tgi}.

So far, we have a choice of coefficients \eqref{goodflat} in \eqref{MellinRR} that gives a putative 1-loop amplitude $M^{R|R}_\text{flat}$ that we know has the correct flat space limit, but which has unphysical poles at $s,t,u=0$ that in principle could be cancelled by subtracting a polynomial divided by $stu$, but in practice are hard to fix using the superconformal Ward identity because of the double sums. We can avoid these unphysical poles by choosing ${c}(m,n,s,t;\sigma,\tau)$, $ {d}(m,s,t;\sigma,\tau)$, and ${ {e}}(m,s,t;\sigma,\tau)$ in \eqref{MellinRR} that are in fact polynomials in $m,n,s,t$ as originally defined, and then demanding that the resulting expression for $M^{R|R}$ matches $M^{R|R}_\text{flat}$ up to the degree 5 polynomial ambiguities ${\bf P}^{(5),i}(s,t;\sigma,\tau)$. The resulting expressions for $c,d,e$ are given in the attached Mathematica result, and are now guaranteed to have both the correct flat space limit and only physical poles. Finally, the coefficients $k_i$ can in principle be fixed using the superconformal Ward identity in terms of just a single coefficient, which corresponds to the single physical degree 4 contact term ambiguity in \eqref{M2222}. In practice this is difficult due to the double sums, so instead we fix most of these coefficients in the next section by demanding a consistent superblock expansion, which is equivalent to imposing the superconformal Ward identity but easier in practice.

\subsection{Extracting CFT data}
\label{CFTData}

We now extract all low-lying CFT data from the $R|R$, i.e. $c_T^{-2}$, and $R|R^4$, i.e. $c_T^{-\frac83}$, correlators using two independent methods. Firstly, we derive an inversion integral formula for each DD in position space, which lets us to extract all CFT data above a certain spin in terms of a single integral, as expected from the Lorentzian inversion formula \cite{Caron-Huot:2017vep}. Secondly, we expand each entire correlator as written in Mellin space in superblocks to extract all CFT data for all spins up to the physical contact term ambiguities that appear in \eqref{M2222}, as well as some unphysical ambiguities for $R|R$ that in principle can be fixed by the superconformal Ward identity. We find that both methods agree in their respective regimes of applicability. We do not extract CFT data from the $R^4|R^4$, i.e. $c_T^{-\frac{10}{3}}$, correlator, since we anyway do not know the $R|D^6R^4$ term that would contribute at the same order, but it would be simple to extract the $R^4|R^4$ data as well from the formulae provided here.

To extract CFT data, we will look at the superblock expansion in the lightcone limit of small $U\sim z$, where conformal blocks are expanded as
\es{lightBlocksExp}{
G_{\Delta,\ell}(U,V)=\sum_{n=0}^\infty U^{\frac{\Delta-\ell}{2}+n}g_{\Delta,\ell}^{[n]}(1-V)\,.
}
Here, the lowest so-called lightcone block in our normalization is
\es{lightconeBlock}{
g_{\Delta,\ell}^{[0]}(1-V)&=\frac{\Gamma(\ell+1/2)}{4^\Delta\sqrt{\pi}\ell!}(1-V)^\ell \,{}_2F_1\left(\frac{\Delta+\ell}{2},\frac{\Delta+\ell}{2},\Delta+\ell,1-V\right)\,,\\
}
and we see that the expansion is naturally organized in terms of twist $t\equiv \Delta-\ell$. Applying this expansion to the superblocks, we observe that blocks in different supermultiplets with the same twist can appear in the same $R$-symmetry channel, so it is convenient to look at channels with the least mixing between different supermultiplets. For instance, the lowest twist long multiplet $(A,0)_{\ell+2,\ell}$ contributes at greater than twist two in the $[0200]$, $[0120]$, and $[0040]$ channels, while all the protected multiplets contribute at lower twists in these channels, so these channels are the simplest for extracting OPE coefficients of protected operators. In particular, if we focus on the lowest twist 2 conformal blocks at $O(U)$, then from \cite{Chester:2014fya} we see that the short superblocks contain the blocks
\es{shortblock}{
&\text{twist = 2:}\qquad\mathfrak{G}_{(B,+)}^{[0040]}=G_{2,0}\,,\qquad \mathfrak{G}_{(B,+)}^{[0120]}=-\frac43G_{3,1}\,,\qquad \mathfrak{G}_{(B,+)}^{[0200]}=0\,,\\
&\qquad\qquad\qquad\;\,\mathfrak{G}_{(B,2)}^{[0040]}=0\,,\qquad\quad\;\, \mathfrak{G}_{(B,2)}^{[0120]}=-\frac83G_{3,1}\,,\qquad \mathfrak{G}_{(B,2)}^{[0200]}=G_{2,0}+\frac{64}{45}G_{4,2}\,,\\
}
while the semishort superblocks contain 
\es{Apblock}{
&\text{twist = 2:}\quad\mathfrak{G}_{(A,+)_\ell}^{[0040]}=\frac{16}{3}G_{\ell+4,\ell+2}\,,\qquad \mathfrak{G}_{(A,+)_\ell}^{[0120]}=-4G_{3+\ell,1+\ell}-\frac{64 (\ell+3)^4G_{5+\ell,3+\ell}}{(2 \ell+5)^2 (2 \ell+7)^2}\,,\\
&\quad\qquad\qquad\;\, \mathfrak{G}_{(A,+)_\ell}^{[0200]}=\frac{32 (\ell+2) (\ell+3)}{3 (2 \ell+3) (2 \ell+7)}G_{4+\ell,2+\ell}\,,\\
&\quad\qquad\qquad\;\,\mathfrak{G}_{(A,2)_\ell}^{[0040]}=0\,,\qquad\qquad\qquad\;\; \mathfrak{G}_{(A,2)_\ell}^{[0120]}=-\frac{32 (\ell+2)^2}{(2 \ell+3) (2 \ell+5)}G_{\ell+4,\ell+2}\,,\\ 
&\quad\qquad\qquad\;\,\mathfrak{G}_{(A,2)_\ell}^{[0200]}=4G_{\ell+3,\ell+1}+\frac{64 (\ell+2)^2 (\ell+3)^2}{(2 \ell+3) (2 \ell+5)^2 (2 \ell+7)}G_{\ell+5,\ell+3}\,.\\
}
For the long superblock, we will only extract its lowest twist 2 anomalous dimension, for which it is convenient to consider the $[0040]$ channel where only a single block at twist 6 appears:
\es{longblock}{
\mathfrak{G}_{\ell+2,\ell}^{[0040]} = \frac{16 (2 \ell+2) (2 \ell+4)}{(2 \ell+3) (2 \ell+5)}G_{\ell+6,\ell}\,.
}
For an anomalous dimension, the coefficient of $G_{\Delta+4,\ell}$ must be kept at general $\Delta$ before expanding:
\[
f(\Delta,\ell)=\frac{128(\Delta-\ell+1)(\Delta-\ell-1)(\Delta+\ell)(\Delta+\ell+2)}{3(\Delta-\ell+2)(\Delta-\ell)(\Delta+\ell+1)(\Delta+\ell+3)}\,.
\]
Equation \eqref{longblock} is its value at $\Delta=\ell+2$.

We can now use this explicit block decomposition to extract CFT data, first using the Lorentzian inversion formula. In Appendix \ref{Lorentz} we review following the similar 4d case in \cite{Alday:2017vkk} how to use the inversion formula to extract CFT data from the DD of a 3d CFT by comparing to its conformal block expansion in the large $c_T$ limit. We can apply this to the twist 2 block expansion in the $[0040]$ channel \eqref{Apblock} for general $\ell$ where only $(A,+)_\ell$ appears, so that we get
\es{ApInversion}{
\lambda^2_{(A,+),\ell} =\frac{12 (2 \ell+5) \Gamma (\ell+3)^4}{\Gamma \left(\ell+\frac{5}{2}\right)^2
   \Gamma \left(\ell+\frac{7}{2}\right)^2} \int_0^1 \frac{d \bar z}{\bar z}    g_{\ell+4,\ell+2}(\bar z) \text{dDisc}[ {\cal G}^{[0040]}(z\bar z,1-\bar z)\vert_z ] \,,
}
where ${\cal G}^{[0040]}(z\bar z,1-\bar z)\vert_z$ denotes the leading $z$ term in the basis \eqref{Ybasis} after crossing, we dropped the superscript from the leading lightcone block in \eqref{lightconeBlock}, and the overall normalization was fixed using the known GFFT term in \eqref{2222data} as discussed in Appendix \ref{Lorentz}. The $(B,+)$ supermultiplet also appears at leading twist in this channel, and comparing \eqref{shortblock} to \eqref{Apblock} we see corresponds to the limit\footnote{In other channels negative spins blocks appear in this limit as discussed in \cite{Chester:2014fya}, so the comparison is more subtle.}
\es{BpfromAp}{
\mathfrak{G}_{(B,+)}^{[0040]}=\frac{3}{16}\lim_{\ell\to-2}\mathfrak{G}_{(A,+)_\ell}^{[0040]}\,,
}
so can be considered a special case of \eqref{ApInversion}. We can similarly analyze the $[0200]$ and $[0120]$ channels to get
\es{0200}{
&\frac{12 (\ell+1)^2 (\ell+2)^2}{(2 \ell+1) (2 \ell+3)^2 (2 \ell+5)} \lambda^2_{(A,2)_{\ell-1}}+
\frac{2 (\ell+2) (\ell+3)}{(2 \ell+3) (2 \ell+7)}\lambda^2_{(A,+)_\ell}+\frac{3}{4}\lambda^2_{(A,2)_{\ell+1}} = \\
&\frac{12 (2 \ell+5) \Gamma (\ell+3)^4}{\Gamma \left(\ell+\frac{5}{2}\right)^2
   \Gamma \left(\ell+\frac{7}{2}\right)^2}  \int_0^1 \frac{d \bar z}{\bar z}    g_{\ell+4,\ell+2}(\bar z) \text{dDisc}[ {\cal G}^{[0200]}(z\bar z,1-\bar z)\vert_z ]  \,,
   }
   and
   \es{0120}{
&\frac{12 (\ell+2)^4}{(2 \ell+3)^2 (2 \ell+5)^2}\lambda^2_{(A,+)_{\ell-1}}+
\frac{6 (\ell+2)^2}{(2 \ell+3) (2 \ell+5)}\lambda^2_{(A,2)_\ell}+\frac{3}{4}\lambda^2_{(A,+)_{\ell+1}} = \\
&-\frac{12 (2 \ell+5) \Gamma (\ell+3)^4}{\Gamma \left(\ell+\frac{5}{2}\right)^2
   \Gamma \left(\ell+\frac{7}{2}\right)^2}  \int_0^1 \frac{d \bar z}{\bar z}    g_{\ell+4,\ell+2}(\bar z) \text{dDisc}[ {\cal G}^{[0120]}(z\bar z,1-\bar z)\vert_z ] \,,
}
where the minus sign is expected because the spin is odd. These two equations give an overconstrained system for $\lambda^2_{(A,2)_\ell}$ in terms of $\lambda^2_{(A,+)_{\ell-1}}$ and $\lambda^2_{(A,+)_{\ell+1}}$, which can be extracted seperately from \eqref{ApInversion}. From comparing \eqref{shortblock} to \eqref{Apblock}, we see that $(B,2)$ corresponds to the limit
\es{B2fromA2}{
\mathfrak{G}_{(B,2)}^{[0200]}=\frac{1}{4}\lim_{\ell\to-1}\mathfrak{G}_{(A,2)_\ell}^{[0200]}\,,
}
and similarly for the $[0040]$ and $[0120]$ channels, so its OPE coefficient can be extracted from $4\lambda^2_{(A,2)_{-1}}$. Finally, we can apply the inversion analysis in Appendix \ref{Lorentz} for the $R|R$ correction to the anomalous dimension to the long superblock in the $[0040]$ channel \eqref{longblock} to get
\es{anom1}{
\gamma^{R|R}_{2,\ell} =& \frac{1}{(\lambda^{(0)}_{2,\ell})^2 }\Big(4 R^{[0040]}_{1,R|R}( \ell) + \frac12 \partial_{\ell} \big[(\lambda^{(0)}_{2,\ell})^2 ( \gamma^{R}_{2,\ell} )^2\big] - (\lambda^{R}_{2,\ell})^2  \gamma^{R}_{2,\ell} \Big)\\
&+\left(A_{2\ell}-\frac{7}{12}\right)(\gamma^R_{2,\ell})^2\,,
}
where $A_n=\sum_{k=1}^n(-1)^{k+1}/k$, with $A_0=0$. The last term accounts for the dimension dependence of the conformal-block normalization and the superblock coefficient. We have the inversion integral
\es{anom2}{
R^{[0040]}_{1,R|R}(\ell) &=\frac{512 (\ell+1) (\ell+2) (2 \ell+3) \Gamma (\ell+1)^4}{\Gamma
   \left(\ell+\frac{1}{2}\right)^2 \Gamma \left(\ell+\frac{5}{2}\right)^2}\int_0^1 {d \bar z}{\bar z}  g_{\ell+6,\ell}(\bar z) \text{dDisc}[\left. {\cal G}_{R|R}^{[0040]}(z\bar z,1-\bar z) \right|_{z^3 \log z}] \,.
}
For $R|R^4$ we have a similar expression, except without the tree level terms in \eqref{anom1} since for $R^4$ they only have support for $\ell=0$.

To apply these inversion integrals, we need to compute the leading $z$ term of the DD in various channels for $R|R$ and $R|R^4$, which is given by the coefficient of $\log^2 (1-\bar z)$ according to
\es{DD}{
{\rm dDisc}\,[ f(z,\bar z) \log^2(1-\bar z) ] = 4\pi^2  f(z,\bar z)\,,
}
for arbitrary $f(z,\bar z)$ analytic at $\bar z=1$. We compute the $\log^2 (1-\bar z)$ term by taking the $U$-slices $h^{(n),[0ab0]}(V)$ in Section \ref{1loopfrom} that multiply $\log^2U$, applying the $1\leftrightarrow3$ crossing \eqref{crossing} to get expressions that multiply $\log^2V\sim\log^2 (1-\bar z)$, resumming the slices, and reexpanding in the $Y_{[0ab,0]}(\sigma,\tau)$ to get the DD in each irrep. For $R|R^4$, we were able to find closed form expressions, while for $R|R$ the result is written in terms of an integral over an auxiliary variable, see the attached \texttt{Mathematica} notebook for the explicit expressions. For $R|R$ we find that the inversion integrals with these explicit DDs converges for $\ell>-\frac12$ for \eqref{ApInversion}, for $\ell>\frac12$ for \eqref{0120}, and for $\ell>\frac32$ for \eqref{anom2}, which allows us to compute the CFT data:
\es{RRfinal}{
(\lambda^{R|R}_{(A,+)_0})^2&=285.32043668331376\ldots\,,\\
(\lambda^{R|R}_{(A,+)_2})^2&=77.098186992813023\ldots\,,\\
(\lambda^{R|R}_{(A,+)_4})^2&=48.536178605208050\ldots\,,\\
(\lambda^{R|R}_{(A,2)_1})^2&=2239.9009500059848\ldots\,,\\
(\lambda^{R|R}_{(A,2)_3})^2&=540.71435539680002\ldots\,,\\
(\lambda^{R|R}_{(A,2)_5})^2&=335.13173417681\ldots\,,\\
\gamma_{2,2}^{R|R}&=\frac{1645242368}{1125 \pi ^4}-\frac{207785984}{663 \pi ^2}\,,\\
\gamma_{2,4}^{R|R}&=\frac{80952501344}{25725 \pi ^4}-\frac{2170015744}{6783 \pi ^2}\,,\\
\gamma_{2,6}^{R|R}&=\frac{390813996864224}{99324225 \pi
   ^4}-\frac{45500125184}{115115 \pi ^2}\,,\\
}
where we could also compute higher spin data if desired.\footnote{From computing some higher spins values, we observed that the anomalous dimensions are not monotonic in spin until $\ell=6$, unlike the case of 4d $\mathcal{N}=4$ SYM \cite{Alday:2018pdi} and 6d $(2,0)$ \cite{Alday:2020tgi} that was monotonic in spin in general. Of course, monotonicity in spin is only required at sufficiently high spin \cite{Komargodski:2012ek}.} Note that the CFT data that we cannot compute, namely $\gamma_{2,0}^{R|R}$, $(\lambda^{R|R}_{(B,+)_0})^2$, and $(\lambda^{R|R}_{(B,2)_0})^2$, is what is affected by the degree 4 contact term $B_4^{R|R}M^{4}$ in \eqref{M2222}, as shown in Table \ref{resultList}, which is analogous to the 4d \cite{Alday:2017xua} and 6d \cite{Alday:2020tgi} cases.  For $R|R^4$ we find that the inversion integrals with the explicit $R|R^4$ DDs converges for $\ell>\frac72$ for \eqref{ApInversion}, for $\ell>\frac92$ for \eqref{0120}, and for $\ell>\frac{11}{2}$ for \eqref{anom2}, which allows us to compute the CFT data:
\es{RR4final}{
(\lambda^{R|R^4}_{(A,+)_4})^2&=\frac{22291954008064 \left(\frac{2}{3}\right)^{2/3}}{4357815 \pi ^{14/3}}+\frac{1561306511441920
   \left(\frac{2}{3}\right)^{2/3}}{6298655363 \pi ^{8/3}}\,,\\
(\lambda^{R|R^4}_{(A,2)_5})^2&=\frac{254814018760343552 \left(\frac{2}{3}\right)^{2/3}}{3277699425 \pi ^{14/3}}+\frac{281474976710656
   \left(\frac{2}{3}\right)^{2/3}}{72177105 \pi ^{8/3}}\,,\\
\gamma_{2,6}^{R|R^4}&=-\frac{512640462848 \left(\frac{2}{3}\right)^{2/3}}{693 \pi ^{14/3}}-\frac{110655386419200\left( \frac23\right)^{2/3} }{2956811 \pi ^{8/3}}\,,\\
}
where we could also compute higher spin data if desired. The CFT data we cannot compute is what is affected by the degree 8 and smaller contact terms in \eqref{M2222}, as shown in Table \ref{resultList}, which is analogous to the 4d \cite{Alday:2018pdi} and 6d \cite{Alday:2020tgi} cases.  

Finally, we can also extract CFT data from the entire correlator as written in Mellin space in terms of the contact term ambiguities described before. We extract this data in the lightcone expansion following \cite{Chester:2018lbz}, by taking the relevant $s$-pole, doing the $t$-integral, projecting against a block of the corresponding spin using the projectors introduced in \cite{Heemskerk:2009pn}, and comparing against the lightcone expansion of the superblocks in \eqref{shortblock}, \eqref{Apblock}, and \eqref{longblock}. For $R|R$, we first demand that $\log U$ terms, which correspond to anomalous dimensions, should only show up in the $[0200]$, $[0120]$, and $[0040]$ channels starting with the appropriate twists, which fixes 12 of the 24 coefficients $k_i$ of the polynomial ambiguity in the Mellin amplitude \eqref{MellinRR}. After fixing these, we find that $\gamma^{R|R}_{2,\ell}$ for $\ell\geq4$, $(\lambda_{(A,+)_\ell})^2$ for $\ell\geq 2$, and $(\lambda_{(A,2)_\ell})^2$ for $\ell\geq3$ are unaffected by the remaining $k_i$, so this data can be extracted by computing the double sums numerically. For the anomalous dimensions, the projection must retain the separate dimension and spin dependence of both the conformal-block normalization and the superblock coefficient $f(\Delta,\ell)$; freezing these factors omits the final term in \eqref{anom1}. For $R|R^4$, we already completely fixed the Mellin amplitude up to the physical contact term ambiguities in \eqref{M2222}, so we can compute all CFT data, which confirms the inversion results in \eqref{RR4final}, and also gives the complete result at order $c_T^{-\frac83}$ for the low spin data:
\es{RR4final2}{
(\lambda^{R|R^4}_{(B,+)})^2&=-\frac{59609927581958144 \left(\frac{2}{3}\right)^{2/3}}{14189175 \pi ^{14/3}}+\frac{256}{35}B_4^{R|R^4}\,,\\
(\lambda^{R|R^4}_{(B,2)})^2&=-\frac{59609927581958144 \left(\frac{2}{3}\right)^{2/3}}{2837835 \pi ^{14/3}}+\frac{256}{7}B_4^{R|R^4}\,,\\
(\lambda^{R|R^4}_{(A,+)_0})^2&=-\frac{7798563930112 \left(\frac{2}{3}\right)^{2/3}}{1216215 \pi ^{14/3}}-\frac{134217728 \left(\frac{2}{3}\right)^{2/3}}{429 \pi ^{8/3}}\\
&\quad+\frac{16384}{1485}B_6^{RR|R^4}+\frac{950272}{6435}B_7^{RR|R^4}-\frac{131396796416}{467137125}B_8^{RR|R^4}\,,\\
(\lambda^{R|R^4}_{(A,+)_2})^2&=-\frac{148820650360832 \left(\frac{2}{3}\right)^{2/3}}{2786875 \pi ^{14/3}}-\frac{229076375699456 \left(\frac{2}{3}\right)^{2/3}}{56581525 \pi ^{8/3}}+\frac{67108864}{557375}B_8^{R|R^4}\,,\\
(\lambda^{R|R^4}_{(A,2)_1})^2&=\frac{3402914332672 \left(\frac{2}{3}\right)^{2/3}}{218295 \pi ^{14/3}}-\frac{794568949760 \left(\frac{2}{3}\right)^{2/3}}{29393 \pi ^{8/3}}\\
&\quad+\frac{131072}{1155}B_6^{R|R^4}+\frac{21889024}{15015}B_7^{R|R^4}-\frac{3848847491072}{1089986625}B_8^{R|R^4}\,,\\
(\lambda^{R|R^4}_{(A,2)_3})^2&=-\frac{3672876448219136 \left(\frac{2}{3}\right)^{2/3}}{2546775 \pi ^{14/3}}-\frac{614077244112896 \left(\frac{2}{3}\right)^{2/3}}{6292363 \pi ^{8/3}}+\frac{268435456}{121275}B_8^{R|R^4}\,,\\
\gamma_{2,0}^{R|R^4}&=\frac{5112797289066496 \left(\frac{2}{3}\right)^{2/3}}{45045 \pi ^{14/3}}+\frac{359760199680 \left(\frac{2}{3}\right)^{2/3} }{46189 \pi ^{8/3}}\\
&\quad-192B_4^{R|R^4}+\frac{15360}{11}B_6^{R|R^4}+\frac{192000}{11}B_7^{R|R^4}-\frac{18059264}{1521}B_8^{R|R^4}\,,\\
\gamma_{2,2}^{R|R^4}&=\frac{12875118112768 \left(\frac{2}{3}\right)^{2/3}}{1365 \pi ^{14/3}}+\frac{162643071467520 \left(\frac{2}{3}\right)^{2/3}  }{96577 \pi ^{8/3}}\\
&\quad-1536B_6^{R|R^4}-18432B7^{R|R^4}+\frac{509591552}{12675}B_8^{R|R^4}\,,\\
\gamma_{2,4}^{R|R^4}&=\frac{1806876913664 \left(\frac{2}{3}\right)^{2/3}}{63 \pi ^{14/3}}+\frac{709927895040\left(\frac{2}{3}\right)^{2/3}  }{391 \pi ^{8/3}}-32768B_8^{R|R^4}\,,\\
}
where note that $(\lambda^{R|R^4}_{(B,2)})^2=5(\lambda^{R|R^4}_{(B,+)})^2$ as expected from \eqref{1drel}.

\section{Fixing 1-loop contact terms}
\label{1loopContact}

So far we computed the 1-loop Mellin amplitudes $M^{R|R}(s,t)$, $M^{R|R^4}(s,t)$, and $M^{R^4|R^4}(s,t)$, but we did not fix the polynomial in $s,t$ contact terms ambiguities that appear in \eqref{M2222} at orders $c_T^{-2}$, $c_T^{-\frac83}$, and $c_T^{-\frac{10}{3}}$, respectively. It is necessary to fix these contact terms if we want to extract low spin CFT data at these orders that are affected by these ambiguities, as summarized by Table \ref{resultList}. We will fix the contact term ambiguities for $M^{R|R}(s,t)$ and $M^{R|R^4}(s,t)$ using two methods. First, we will propose a unique analytic continuation of Lorentzian inversion below the spins where it was shown to converge in the previous section, which will allow us to fix all CFT data at order $c_T^{-2}$, and all but the CFT data affected by the $B^{R|R^4}_4M^4$ contact term at order $c_T^{-\frac83}$. We will then use the two localization constraints from \cite{Chester:2018aca,Binder:2018yvd} to confirm these results at order $c_T^{-2}$, as well as fix $B^{R|R^4}_4$ and give an additional nontrivial consistency check. Note that at order $c_T^{-2}$, we will be able to extract all CFT data even though we will not write down an explicit Mellin amplitude, because we have not yet fixed all the coefficients $k_i$ in \eqref{MellinRR} that are in principle fixed by the superconformal Ward identity. Also, we cannot similarly analyze the order $c_T^{-\frac{10}{3}}$ term yet, because it receives contributions not only from $M^{R^4|R^4}(s,t)$ but also from $M^{R|D^6R^4}(s,t)$, which has not yet been computed.

\subsection{Analytic continuation of Lorentzian inversion}
\label{analyticCon}

Let us begin by discussing the inversion integral \eqref{ApInversion} for the $[0040]$ channel, which we use to compute $\lambda^2_{(A,+)_\ell}$ for even $\ell$. For $R|R$, we can see from the explicit expression for the DD in the attached \texttt{Mathematica} file that it has the small $\bar z$ expansion
\es{smallzb}{
\text{dDisc}[ {\cal G}_{R|R}^{[0040]}(z\bar z,1-\bar z)\vert_z ]=\frac{3072}{\pi  {\bar z}^{3/2}}-\frac{61952}{27 \pi  \sqrt{\bar z}}+\dots\,.
}
Since the measure in \eqref{ApInversion} scales as ${\bar z}^{\ell+1}$, we see that this integral converges for $\ell>-\frac12$, which allowed us to compute all the $(\lambda^{R|R}_{(A,+)_\ell})^2$ for even $\ell\geq0$ in the previous section, but did not allow us to compute $(\lambda^{R|R}_{(B,+)})^2=\frac{16}{3}(\lambda^{R|R}_{(A,+)_{-2}})^2$ as given by \eqref{BpfromAp}. We can uniquely analytically continue \eqref{ApInversion} to $\ell=-2$ by writing it as
\es{ApInversion2}{
\lambda^2_{(A,+),\ell} &=\frac{12 (2 \ell+5) \Gamma (\ell+3)^4}{\Gamma \left(\ell+\frac{5}{2}\right)^2
   \Gamma \left(\ell+\frac{7}{2}\right)^2} \Bigg[\int_0^1 \frac{d \bar z}{\bar z}    g_{\ell+4,\ell+2}(\bar z)\Big[ \text{dDisc}[ {\cal G}^{[0040]}(z\bar z,1-\bar z)\vert_z ]\\
   &\quad-\frac{3072}{\pi} \left(\frac{1-\bar z}{\bar z}\right)^{3/2}-\frac{62464}{27 \pi } \sqrt{\frac{1-z}{z}} \Big]+\frac{3072}{\pi}{\bf f}(\ell,3/2)+\frac{62464}{27 \pi } {\bf f}(\ell,1/2)\Bigg]\,,
}
where we define the analytically continued integral
\es{analInt}{
\int_0^1 \frac{d \bar z}{\bar z}    g_{\ell+4,\ell+2}(\bar z)\left(\frac{1-\bar z}{\bar z}\right)^{p}=\frac{\Gamma (2 (\ell+3)) \Gamma (p+1)^2 \Gamma (\ell-p+2)}{\Gamma (\ell+3)^2 \Gamma (\ell+p+4)}\,,
}
which can be computed using the integral expression for the hypergeometric function in the lightcone block \eqref{lightconeBlock}. The explicit $\bar z$ integral in \eqref{ApInversion2} as well as $\bf{f}(\ell,3/2)$ and $\bf{f}(\ell,1/2)$ are now convergent for $\ell<-\frac12$, which we can compute at high precision for $\ell=-2$ to get
\es{BpRRfinal}{
(\lambda^{R|R}_{(B,+)})^2=\frac{16}{3}(\lambda^{R|R}_{(A,+)_{-2}})^2=\frac{32768}{45 \pi ^2}-\frac{81920}{9 \pi ^4}\,.
}
We can similarly analytically continue the inversion integrals in the $[0120]$ \eqref{0120} and $[0200]$ \eqref{0200} channels to compute $(\lambda^{R|R}_{(B,2)})^2=4(\lambda^{R|R}_{(A,2)_{-1}})^2$, which is related to $(\lambda^{R|R}_{(B,+)})^2$ by a factor of 5 as expected from \eqref{1drel}, which is evidence that our analytic continuation respects superconformal symmetry. The analytic continuation of the inversion integral \eqref{anom2} for the anomalous dimension then gives
\es{A0RRfinal}{
\gamma_{2,0}^{R|R}=\frac{39639040}{9 \pi ^4}-\frac{117698560}{429 \pi ^2}\,,
}
which is the only other CFT data that was affected by the $B_4^{R|R}M^4(s,t)$ contact term, and so could not be computed in the previous section.

We can then analytically continue the inversion integrals in the same way for $R|R^4$. In this case we find that the small $\bar z$ expansion of the DD includes a term ${\bar z}^{-1}$, which gives a ${\bf{f}}(\ell,1)$ term after analytic continuation. From \eqref{analInt}, we see that this ${\bf{f}}(\ell,1)$ has a logarithmic divergence at $\ell=-2$, so we can only analytically continue for $\ell>-2$, which allows us to compute all $(\lambda^{R|R}_{(A,+)_\ell})^2$ for even $\ell\geq0$, but not $(\lambda^{R|R^4}_{(B,+)})^2=\frac{16}{3}(\lambda^{R|R^4}_{(A,+)_{-2}})^2$. We see a similar pattern in the other inversion integrals, where we can compute all CFT except $(\lambda^{R|R^4}_{(B,2)})^2$ and all $\gamma^{R|R^4}_{2,0}$, which would be affected by the $B_4^{R|R^4}M^4(s,t)$ contact term. The results for the other CFT data are\footnote{Curiously, the results for the anomalous dimensions for $R|R^4$ become monotonic in spin at $\ell=6$, which was the same value for $R|R$ as discussed above.}
\es{RR4final3}{
(\lambda^{R|R^4}_{(A,+)_0})^2&=-\frac{269877248 \left(\frac{2}{3}\right)^{2/3}}{45 \pi ^{14/3}}-\frac{134217728 \left(\frac{2}{3}\right)^{2/3}}{429 \pi ^{8/3}}\,,\\
(\lambda^{R|R^4}_{(A,+)_2})^2&=-\frac{7322684358656 \left(\frac{2}{3}\right)^{2/3}}{91875 \pi ^{14/3}}-\frac{229076375699456 \left(\frac{2}{3}\right)^{2/3}}{56581525
   \pi ^{8/3}}\,,\\
(\lambda^{R|R^4}_{(A,2)_1})^2&=-\frac{167914766336 \left(\frac{2}{3}\right)^{2/3}}{315 \pi ^{14/3}}-\frac{794568949760 \left(\frac{2}{3}\right)^{2/3}}{29393 \pi
   ^{8/3}}\,,\\
(\lambda^{R|R^4}_{(A,2)_3})^2&=-\frac{1634776490442752 \left(\frac{2}{3}\right)^{2/3}}{848925 \pi ^{14/3}}-\frac{614077244112896
   \left(\frac{2}{3}\right)^{2/3}}{6292363 \pi ^{8/3}}\,,\\
   \gamma_{2,0}^{R|R^4}&=\frac{25509449728 \left(\frac{2}{3}\right)^{2/3}}{15 \pi ^{14/3}}+\frac{359760199680 \left(\frac{2}{3}\right)^{2/3}}{46189 \pi ^{8/3}}\,,\\
\gamma_{2,2}^{R|R^4}&=\frac{166203518976 \left(\frac{2}{3}\right)^{2/3}}{5 \pi ^{14/3}}+\frac{162643071467520 \left(\frac{2}{3}\right)^{2/3}}{96577 \pi ^{8/3}}\,,\\
\gamma_{2,4}^{R|R^4}&=\frac{2257848479744 \left(\frac{2}{3}\right)^{2/3}}{63 \pi ^{14/3}}+\frac{709927895040 \left(\frac{2}{3}\right)^{2/3}}{391 \pi ^{8/3}}\,.\\
}
We can compare this against the CFT data in \eqref{RR4final2} that was extracted from the explicit Mellin amplitude, which fixes the coefficients
\es{fixB}{
B_6^{R|R^4}=-\frac{128720195584 \left(\frac{2}{3}\right)^{2/3}}{819 \pi ^{14/3}}\,,\qquad B_7^{R|R^4}=\frac{775420813312 \left(\frac{2}{3}\right)^{2/3}}{68445 \pi ^{14/3}}\,,\qquad B_8^{R|R^4}=-\frac{655360 \left(\frac{2}{3}\right)^{2/3}}{3 \pi ^{14/3}}\,.
}
We cannot yet fix the $B_4^{R|R^4}$ coefficient, because the analytic continuation of inversion did not converge for the low spin data affected by $M^4(s,t)$.

\subsection{Supersymmetric localization}
\label{loc}

We can also fix the Mellin amplitudes using the two localization constraints in \cite{Chester:2018aca,Binder:2018yvd}. The first constraint is simply the value of the short OPE coefficients $\lambda^2_{(B,+)}$ and $\lambda^2_{(B,2)}$, which are shown to all orders in $1/c_T$ in \eqref{B2Bp}, and impose just one independent constraint on the 4-point function due to the relation \eqref{1drel}. For $R|R$, the localization values exactly match the prediction in \eqref{BpRRfinal} from the analytically continued Lorentzian inversion formula, which independently fixes the CFT data affected by $M^4(s,t)$ without needing to assume the conjectured analytic continuation, and so gives a nontrivial check on that conjecture. For $R|R^4$ we use this constraint to fix the last contact term ambiguity
\es{lastB}{
B_4^{R|R^4} = \frac{65229926487808 \left(\frac{2}{3}\right)^{2/3}}{135135 \pi ^{14/3}}\,,
}
which we can then use to compute the last remaining unfixed CFT datum
\es{lastAnom}{
\gamma^{R|R^4}_{2,0}=\frac{25509449728 \left(\frac{2}{3}\right)^{2/3}}{15 \pi ^{14/3}}+\frac{359760199680 \left(\frac{2}{3}\right)^{2/3}}{46189 \pi ^{8/3}}\,.
}

The second localization constraint involves a nontrivial integral \cite{Binder:2019mpb}:
 \es{DerSimpTwoMassesFinal}{
&\frac{\partial \log Z}{\partial m_+^2 \partial m_-^2} \Big\vert_{m_\pm=0}= \frac{\pi^2 c_T^2}{2^{11}}I_{+-}[\cS^i]\,,\\
&\qquad\quad\;\;\, I_{+-}[{\cal S}^i]  \equiv \int \frac{ds\ dt}{(4\pi i)^2} \frac{2\sqrt{\pi}}{(2-t)(s+t-2)}\mathfrak{M}_1(s,t) \\
&\qquad\qquad\quad\times \Gamma \left[1-\frac{s}{2}\right] \Gamma \left[\frac{s+1}{2}\right] \Gamma \left[1-\frac{t}{2}\right] \Gamma \left[\frac{t-1}{2}\right] \Gamma \left[\frac {s+t-2}{2}\right] \Gamma \left[\frac{3-s-t}2\right]\,,
}
where $\mathfrak{M}_1(s,t)$ is the first element of the Mellin amplitude basis
\es{Mbasis}{
M(s,t;\sigma,\tau) = \mathfrak{M}_1+\sigma^2 \mathfrak{M}_2+\tau^2 \mathfrak{M}_3+\sigma\tau \mathfrak{M}_4+\tau  \mathfrak{M}_5+\sigma  \mathfrak{M}_6\,,
}
and the mass derivatives of the partition function was computed to all orders in $1/c_T$ in \cite{Binder:2018yvd} for $k=2$ ABJ(M):
\es{Z}{
\frac{1}{c_T^2}\frac{\partial \log Z}{\partial m_+^2 \partial m_-^2} \Big\vert_{m_\pm=0}=-\frac{\pi ^2}{64
  c_T}+\frac{5 \pi ^{4/3} \left(\frac23\right)^{\frac23}}{16c_T^{5/3}}-\frac{5}{12
  c_T^2}-\frac{4 \left(\frac{2\pi^2}{3}\right)^{\frac13}}{3c_T^{7/3}}+\frac{91
   \left(\frac{2}{3 \pi }\right)^{2/3}}{9c_T^{8/3}}+O(c_T^{-3})\,.
}
For the Mellin amplitudes that appear at order $c_T^{-\frac83}$ in \eqref{M2222} we compute the integrals in \eqref{DerSimpTwoMassesFinal} to get
\es{ints}{
I_{+-}[M^4] &= \frac{8 \pi ^2}{7}\,,\qquad I_{+-}[M^6] =-\frac{448 \pi ^2}{33}\,,\qquad I_{+-}[M^7] = -\frac{529472 \pi ^2}{3003}\,,\\
 I_{+-}[M^8] &= \frac{49716397568 \pi ^2}{342567225}\,,\qquad  I_{+-}[M^{R|R^4}]= -\frac{1861955980828672 \left(\frac{2}{3}\right)^{2/3}}{2837835 \pi ^{8/3}}\,,
}
where for the polynomial Mellin amplitudes we used the fact that they are all proportional to $(2-t)(s+t-2)$ as well as the Barnes Lemma
\es{barnes}{
\int_{-i\infty}^{i\infty}\frac{ds}{2\pi i}\Gamma(a+s)\Gamma(b+s)\Gamma(c-s)\Gamma(d-s) = \frac{\Gamma(a+c)\Gamma(b+d)\Gamma(b+c)\Gamma(a+d)}{\Gamma(a+b+c+d)} \,,
}
while for $M^{R|R^4}(s,t)$ we instead numerically computed the integral of the closed form expression in the \texttt{Mathematica} file to high precision. Plugging \eqref{ints} and \eqref{Z} into \eqref{DerSimpTwoMassesFinal}, we find that $M^{R|R^4}$ with the values of $B_i^{R|R^4}$ fixed in \eqref{lastB} and \eqref{fixB} precisely satisfies the constraint, which is a nontrivial check that the analytically continued Lorentzian inversion formula gives the correct CFT data and thus fixes the contact term ambiguities at 1-loop.

Finally, we can use both localization constraints and the explicit integrals in \eqref{ints} to fix two of the four coefficients $B_i^{D^8R^4}$ in the tree level $D^8R^4$ term in \eqref{M2222} to get
\es{D8R4answer}{
B^{D^8R^4}_4=-\frac{3200B^{D^8R^4}_7 }{429}-\frac{238578176B^{D^8R^4}_8}{1957527}\,,\quad B^{D^8R^4}_6=-\frac{177B^{D^8R^4}_7}{13}+\frac{10356296B^{D^8R^4}_8 }{494325}\,,
} 
which will be useful in future attempts to fix this amplitude by independently computing its CFT data.

\section{Numerical bootstrap}
\label{numerics}

In the previous sections we studied $\langle2222\rangle$ for the $k=2$ ABJ(M) theory in the large $c_T\sim N^{\frac32}$ limit to several orders. In this section, we will study this correlator non-perturbatively using the numerical conformal bootstrap, and compute bounds on CFT data as a function of $c_T$ as was done in previous work \cite{Chester:2014fya,Chester:2014mea,Agmon:2017xes,Agmon:2019imm}, but now to much higher numerical accuracy using subsequent technical improvements to the bootstrap software \cite{Landry:2019qug}. Previously, the tree level supergravity correction \cite{Zhou:2017zaw,Chester:2018lbz} was found to saturate the lower bounds \cite{Chester:2014mea,Agmon:2017xes} for all CFT. These lower bounds were conjectured to correspond to $k=2$ ABJ(M) theory in \cite{Agmon:2017xes}, because they were found to be approximately saturated by the values of the short $(B,2)$ and $(B,+)$ OPE coefficients as computed all orders in $1/N$ in \cite{Agmon:2017xes}. We now find that the $R|R$ correction continues to saturate these lower bounds for those semishort $(A,2)_\ell$ and $(A,+)_\ell$ OPE coefficients where the asymptotic large $c_T$ expansion is well converged. 

\subsection{Setup}
\label{setupNum}

We start by briefly reviewing how the numerical bootstrap can be applied to the stress-tensor multiplet four-point function in $\mathcal{N}=8$ theories, for further details see \cite{Chester:2014fya}. Invariance of the four-point function \eqref{4point} as expanded in superblocks \eqref{SBDecomp} under $1\leftrightarrow3$ crossing \eqref{crossing} implies crossing equations of the form
 \es{crossingEq}{
 \sum_{{\cal M}\, \in\,\{\text{Id},\,\text{Stress},\,(B,+),\,(B,2),\,(A,+)_\ell,\,(A,2)_\ell,\,(A,0)_{\Delta,\ell}\} } \lambda_{\cal M}^2\,  \vec{V}_{{\cal M}} = 0 \,,
 }
where ${\cal M}$ ranges over all the superconformal multiplets listed in Table~\ref{opemult}, $\vec{V}_{{\cal M}} $ are functions of superconformal blocks, and $\lambda_{\cal M}^2$ are squares of OPE coefficients that must be positive by unitarity.  As in \cite{Chester:2014fya}, we normalize the OPE coefficient of the identity multiplet to $\lambda_{\text{Id}} = 1$, and parameterize our theories by the value of $\lambda_\text{Stress}$, which is related to $c_T$ through \eqref{cTolam} for $p=2$.

To find upper/lower bounds on a given OPE coefficient of a protected multiplet ${\cal M}'$ that appears in the ${\cal O}_{\text{Stress}} \times {\cal O}_{\text{Stress}}$ OPE, we consider linear functionals $\alpha$ satisfying
 \es{CondOPE}{
  &\alpha(\vec{V}_{\cal M'}) = s \,, \;\quad\qquad  \text{$s=1$ for upper bounds, $s=-1$ for lower bounds} \,,  \\
  &\alpha(\vec{V}_{\cal M}) \geq 0 \,, \;\;\quad\qquad \text{for all short and semi-short ${\cal M} \notin \{ \text{Id}, \text{Stress}, \cal M' \}$} \,, \\
  &\alpha(\vec{V}_{(A,0)_{\Delta,\ell}}) \geq 0 \,, \qquad \text{for all $\ell$ with $\Delta\geq \ell+1$} \,.\\
 }
If such a functional $\alpha$ exists, then this $\alpha$ applied to \eqref{crossingEq} along with the positivity of all $\lambda_{\cal M}^2$ except, possibly, for that of $\lambda_{{\cal M}'}^2$ implies that
 \es{UpperOPE}{
  &\text{if $s=1$, then}\qquad \lambda_{{\cal M}'}^2 \leq - \alpha (\vec{V}_\text{Id})  -\frac{256}{c_T}\alpha( \vec{V}_\text{Stress} ) \,,\\
    &\text{if $s=-1$, then}\qquad \lambda_{{\cal M}'}^2 \geq  \alpha (\vec{V}_\text{Id})  + \frac{256}{c_T} \alpha( \vec{V}_\text{Stress} ) \,.\\
 }
 Note that we can get both upper/lower bounds because the protected multiplets are isolated from the continuum of operators, unlike the long multiplets $(A,0)_{\Delta,\ell}$ for which we could only compute upper bounds on their OPE coefficients. To obtain the most stringent upper/lower bound on $\lambda_{{\cal M}'}^2$, one should then minimize/maximize the RHS of \eqref{UpperOPE} under the constraints \eqref{CondOPE}. In the above algorithms, we fixed the SCFT by inputting the value of $c_T$, which was computed to all orders in $1/N$ for ABJ(M) in \cite{Agmon:2017xes}. We can further fix the theory by also putting in the values of all the short OPE coefficients $\lambda^2_{(B,2)}$ and $\lambda^2_{(B,+)}$, which were also computed to all orders in $1/N$. We should then remove these operators from the second line of \eqref{CondOPE} and put them on the RHS of \eqref{UpperOPE} with their explicit OPE coefficients, just like the stress tensor multiplet.

The numerical implementation of the minimization/maximization problem described depends on five parameters: the number of derivatives parameter $\Lambda$ used to construct $\alpha$, the range of spins of multiplets up to $\ell_\text{max}$ that we consider, the order $r_\text{max}$ to which we expand blocks, the parameter $\kappa$ that parametrizes how many poles we keep when approximating blocks, and the precision of the solver {\tt SDPB} \cite{Simmons-Duffin:2015qma}. We used $\Lambda=83$,\footnote{This is equivalent to $n_\text{max}=42$, which must be some kind of record for 3d bootstrap!} $\ell_\text{max}=90$, $r_\text{max}=140$, $\kappa=70$, and $1116$ binary digits of precision. The most important parameter is $\Lambda$, which should be compared to the previously highest value $\Lambda=43$ used in \cite{Agmon:2019imm}.

\subsection{Bootstrap bound saturation}
\label{saturation}

\begin{table}[]
\begin{center}
\begin{tabular}{c||c|c|c}
 & $\frac{\lambda_\text{Stress}^2}{16}=\frac{16}{c_T}$& $\lambda_{(B,+)}^2$ & $\lambda_{(B,2)}^2$\\
 \hline \hline
 large $N$    & $0.20952$  & $ 7.36854$  &   $7.43343$ \\
exact   &   $0.20944$   &$7.37115$  &  $7.45176$ \\
\hline
\end{tabular}
\caption{Comparison of the large $N$ formulae to the exact values from \cite{Agmon:2019imm} for the OPE coefficients of short operators that appear in $\langle2222\rangle$ for $U(3)_2\times U(3)_{-2}$ ABJM.\label{compTab}}
\end{center}
\end{table}

\begin{figure}[]
\begin{center}
   \includegraphics[width=0.85\textwidth]{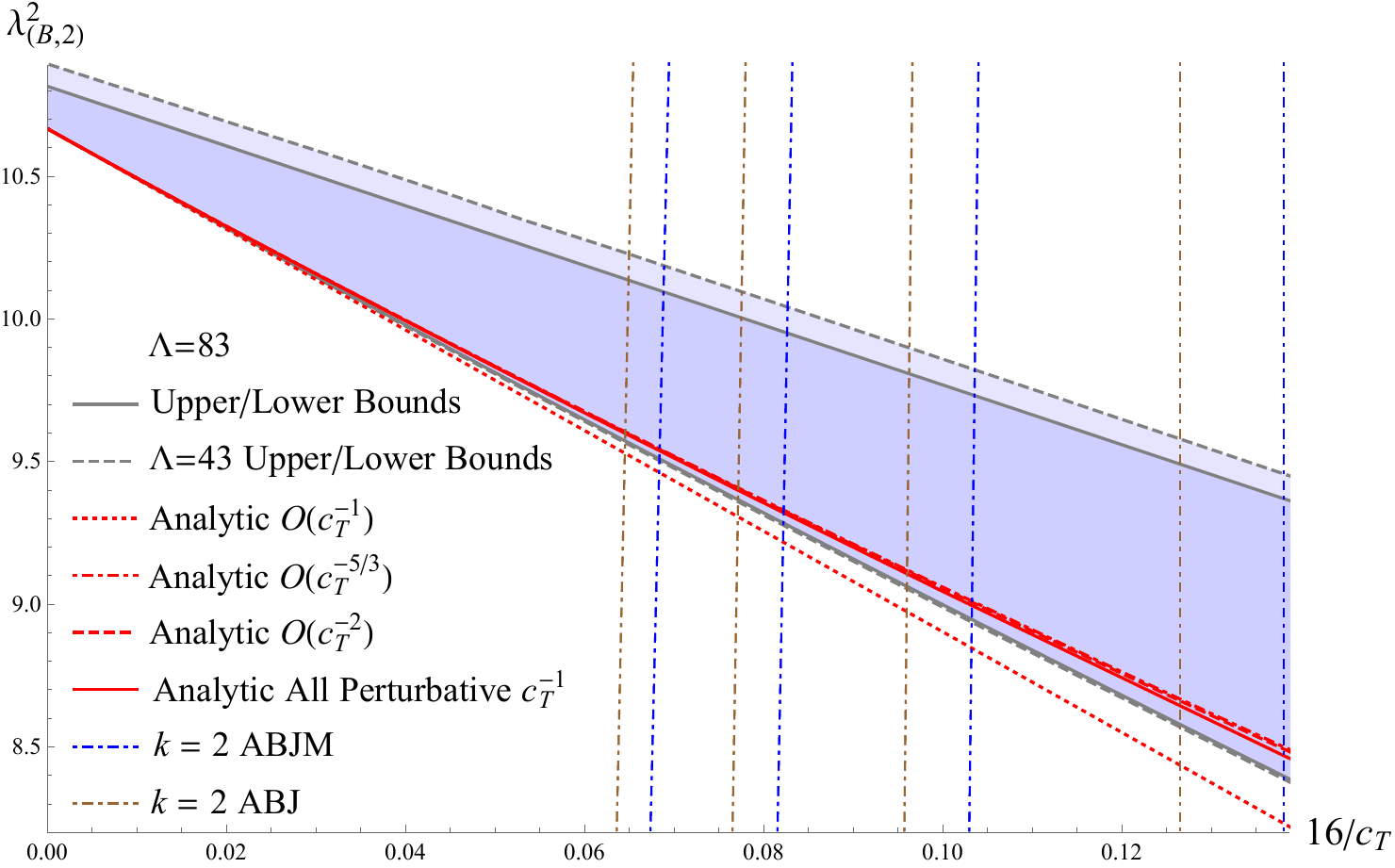}
 \caption{Upper and lower bounds on the $\lambda_{(B,2)}^2$ OPE coefficient with $\Lambda=83$ (solid gray) and $\Lambda=43$ (dashed gray) in terms of the stress-tensor coefficient $c_T$ in the large $c_T$ regime. The red dotted line denotes the large $c_T$ expansion to order tree level supergravity $O(c_T^{-1})$, which is independent of $k$, while the red dot-dashed line also includes the tree level $R^4$ correction at order $O(c_T^{-5/3})$ and the red dashed line furthermore includes the 1-loop $R|R$ correction at order $O(c_T^{-2})$, both of which depend on the value $k=2$. The solid red line includes the all orders in $1/c_T$ expression, which only misses non-perturbative in $c_T$ corrections. The blue and brown vertical lines denote the values of $c_T$ for various known $k=2$ ABJM and ABJ theories, respectively, which are summarized in Table \ref{cTValues}. Since there is still a small discrepancy in this very zoomed in plot (relative to \cite{Agmon:2017xes}, which looked at $0\leq16/c_T\leq1$) between the all orders result and the numerical bounds even at very high $\Lambda$, we have imposed the value of $\lambda^2_{(B,2)}$ in subsequent plots, which should further constrain the numerical bounds to match $k=2$ ABJM.}
\label{B2fig}
\end{center}
\end{figure}

The large $c_T$ expansion of CFT data is asymptotic, which means that after a few orders the expansion will actually get worse, unless we look at very large values of $c_T$. The larger the value of $c_T$, the more precise our numerics must be to make a meaningful comparison, so we should focus on CFT data for which the asymptotic expansion is still pretty good for the lowest few orders. In general, we observe that the convergence of the asymptotic expansion is better for more protected operators. For instance, the short operators $(B,2)$ and $(B,+)$ are the most protected, and their expansion at large $c_T$ as shown to all orders in \cite{Agmon:2017xes} takes the form
\es{B2num}{
\lambda^2_{(B,2)}=10.6667 - 17.6369 \frac{16}{c_T} + 7.26668 \Big[\frac{16}{c_T}\Big]^{\frac53} - 0.384051 \Big[\frac{16}{c_T}\Big]^2 - 
 3.25726 \Big[\frac{16}{c_T}\Big]^{\frac73} + 1.88158 \Big[\frac{16}{c_T}\Big]^{\frac{8}{3}}+\dots\,,
}
and similarly for $\lambda^2_{(B,+)}$. We expand in $16/c_T$ because the free $\mathcal{N}=8$ theory has $c_T=16$, which makes this a natural quantity. Note that coefficient of each subsequent order is in general getting smaller, which implies that this asymptotic expansion is expected to be pretty good even to many orders. This expectation is supported by the fact that the all orders in $1/N$ expansion matches the finite $N$ values to the sub-percent level for the $U(3)_2\times U(3)_{-2}$ ABJM theory, as computed in \cite{Agmon:2019imm}\footnote{Actually, in  \cite{Agmon:2019imm} the values were computed directly for the interacting sector of $U(4)_{1}\times U(4)_{-1}$ ABJM theory, but this theory is dual to $U(3)_2\times U(3)_{-2}$ theory since in the UV they are both described by $SU(4)\cong SO(6)$ SYM \cite{Gang:2011xp,Agmon:2017lga}.} and reviewed in Table \ref{compTab}. Another piece of evidence is that the all orders expression is close to saturating the lower bound from the numerical bootstrap, as first observed in \cite{Agmon:2017xes} with $\Lambda=43$ accuracy, and now further confirmed with $\Lambda=83$ accuracy in Figure \ref{B2fig}. Note that there is still a discrepancy between the all orders expression in solid red and the lower bound in solid gray, even though the numerics seem well converged as can be seen from comparing to the old $\Lambda=43$ value in dashed gray. This suggests that either the lower bound is not actually saturated by the $k=2$ ABJ(M) theory, even though it is very close to the curve for a large range of $c_T$ as observed in \cite{Agmon:2017xes}, or that the non-perturbative in $c_T$ corrections to the all orders expressions for $\lambda^2_{(B,2)}$ might account for this small discrepancy. Since we cannot rule out either possibility at this stage, for subsequent plots we will show bounds where we inputed the values of $\lambda^2_{(B,2)}$ and $\lambda^2_{(B,+)}$ as well as bounds with no assumptions, and as expected these bounds differ by a small amount.

The next most protected operators are $(A,+)_\ell$ for even $\ell$ and $(A,2)_\ell$ for odd $\ell$. In the previous sections we computed the large $c_T$ expansion of their OPE coefficients to many orders, which we summarize showing explicit numerical values for each term:
\es{ApNum}{
\lambda^2_{(A,+)_{0}}&=7.11111 + 3.02803 \frac{16}{c_T} + 1.11453 \Big[\frac{16}{c_T}\Big]^2 - 3.04011 \Big[\frac{16}{c_T}\Big]^{\frac73} \\
&+\Big[\frac{9199616}{6435}B_7^{D^8R^4}-\frac{120501022588928}{5138508375}B_8^{D^8R^4}\Big]{c_T^{-\frac{23}{9}}}- 20.4134 \Big[\frac{16}{c_T}\Big]^{\frac83}+\dots\,,\\
\lambda^2_{(A,+)_{2}}&=13.3747 + 3.19665  \frac{16}{c_T} + 0.301165  \Big[\frac{16}{c_T}\Big]^2+\frac{67108864}{557375}B_8^{D^8R^4}{c_T^{-\frac{23}{9}}} - 268.868  \Big[\frac{16}{c_T}\Big]^{\frac83}+\dots\,,\\
\lambda^2_{(A,+)_{4}}&=19.6506 + 3.25967  \frac{16}{c_T} + 0.189594  \Big[\frac{16}{c_T}\Big]^2 + 16.9909  \Big[\frac{16}{c_T}\Big]^{\frac83}+\dots\,,\\
}
and
\es{A2Num}{
\lambda^2_{(A,2)_{1}}&=9.75238 - 6.17281 \frac{16}{c_T} + 8.74961 \Big[\frac{16}{c_T}\Big]^2  - 75.0472 \Big[\frac{16}{c_T}\Big]^{\frac73} \\
&-\Big[ \frac{262144}{3003}B_7^{D^8R^4}+\frac{2766298677248}{2397970575}B_8^{D^8R^4} \Big]{c_T^{-\frac{23}{9}}}- 1797.23 \Big[\frac{16}{c_T}\Big]^{\frac83}+\dots\,,\\
\lambda^2_{(A,2)_{3}}&=16.2118 - 6.43488 \frac{16}{c_T} + 2.11217 \Big[\frac{16}{c_T}\Big]^2 +\frac{268435456}{121275}B_8^{D^8R^4}{c_T^{-\frac{23}{9}}} - 6491.08 \Big[\frac{16}{c_T}\Big]^{\frac83}+\dots\,,\\
\lambda^2_{(A,2)_{5}}&=22.573 - 6.54041 \frac{16}{c_T} + 1.30911 \Big[\frac{16}{c_T}\Big]^2  + 261.161 \Big[\frac{16}{c_T}\Big]^{\frac83}+\dots\,,\\
}
where we included the $D^8R^4$ at order $c_T^{-\frac{23}{9}}$ term that we only know up to two as yet unfixed coefficients. As the spin increases, we observe that the $R|R$ term at order $c_T^{-2}$ becomes increasingly smaller compared to the tree level supergravity term, which means that we can trust it more for a larger range of $c_T$. On the other hand, the $R|R^4$ term at order $c_T^{-\frac83}$ is much bigger than the previous terms, which means that we can only trust it at very large $c_T$. There is also a $D^6R^4$ term at order $c_T^{-\frac73}$ that only affects the lowest spin for each multiplet, and also is roughly the same size as the supergravity term. 

Lastly, the least protected multiplet is the long multiplet $(A,0)_{\Delta,\ell}$. In the previous sections we computed the large $c_T$ expansion to the scaling dimension of the lowest twist operator for the lowest few spins, which we  summarize showing explicit numerical values for each term:
\es{A0Num}{
\Delta_{2,0}&=2 - 7.09248 \frac{16}{c_T} - 38.1501 \Big[\frac{16}{c_T}\Big]^{\frac53} + 68.034 \Big[\frac{16}{c_T}\Big]^2  + 21.3758 \Big[\frac{16}{c_T}\Big]^{\frac73}\\
& +\Big[-\frac{222720}{143}B_7^{D^8R^4}+\frac{18902167552}{1087515}B_8^{D^8R^4}\Big]{c_T^{-\frac{23}{9}}}+ 
 3993.9 \Big[\frac{16}{c_T}\Big]^{\frac83}+\dots\,,\\
\Delta_{4,2}&=4 - 3.12069 \frac{16}{c_T} - 65.3944 \Big[\frac{16}{c_T}\Big]^2\\
& +\Big[\frac{32256}{13}B_7^{D^8R^4}+\frac{1322266624}{164775}B_8^{D^8R^4}\Big]{c_T^{-\frac{23}{9}}} + 112035. \Big[\frac{16}{c_T}\Big]^{\frac83}+\dots\,,\\
\Delta_{6,4}&=6 - 2.06985 \frac{16}{c_T} - 0.42667 \Big[\frac{16}{c_T}\Big]^2 -32768B_8^{D^8R^4}{c_T^{-\frac{23}{9}}} + 120791. \Big[\frac{16}{c_T}\Big]^{\frac83}+\dots\,.\\
}
The asymptotic expansion for this quantity seems very poor, as the coefficient of each subsequent term is growing rapidly.

\begin{table}%
\begin{center}
\begin{tabular}{|l|c|c|}
\hline
 \multicolumn{1}{|c|}{${\cal N} = 8$ SCFT} & $c_T$ & $\frac{\lambda_\text{Stress}^2}{16}=\frac{16}{c_T}$ \\
  \hline
    $\;\; U(4)_2 \times U(4)_{-2}$  \; ABJM & $126.492$ & $0.138133$\\  
       $\;\; U(5)_2 \times U(5)_{-2}$  \; ABJM & $172.058$ & $0.0998481$\\  
          $\;\; U(6)_2 \times U(6)_{-2}$  \; ABJM & $221.97$ & $0.0765165$\\  
             $\;\; U(7)_2 \times U(7)_{-2}$  \; ABJM & $275.879$ & $0.0610587$\\  
  $\;\; U(4)_2 \times U(5)_{-2}$ \; ABJ  & $115.831$ & $0.12649$\\
   $\;\; U(5)_2 \times U(6)_{-2}$ \; ABJ  & $160.243$ & $0.0929919$\\
    $\;\; U(6)_2 \times U(7)_{-2}$ \; ABJ  & $209.105$ & $0.0720818$\\
     $\;\; U(7)_2 \times U(8)_{-2}$ \; ABJ  & $262.043$ & $0.0579965$\\
   \hline
\end{tabular}
\end{center}
\caption{Several values of $c_T$ and $\lambda_\text{Stress}^2/16$ for $k=2$ ABJM or ABJ theories, as computed from the all orders in $1/N$ formulae in \cite{Agmon:2017xes}.}\label{cTValues}
\end{table}%

\begin{figure}[]
\begin{center}
   \includegraphics[width=0.85\textwidth]{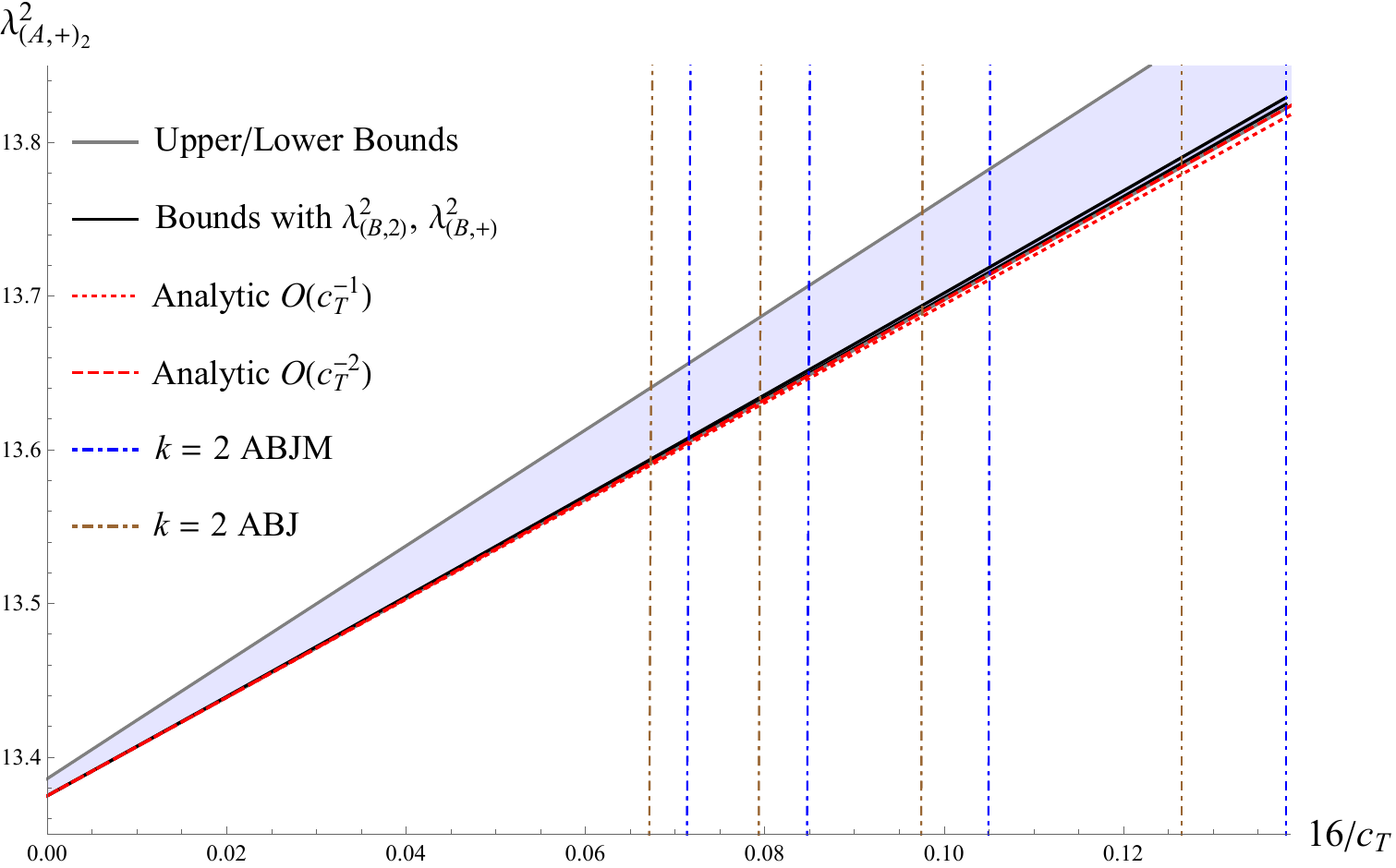}
     \includegraphics[width=0.85\textwidth]{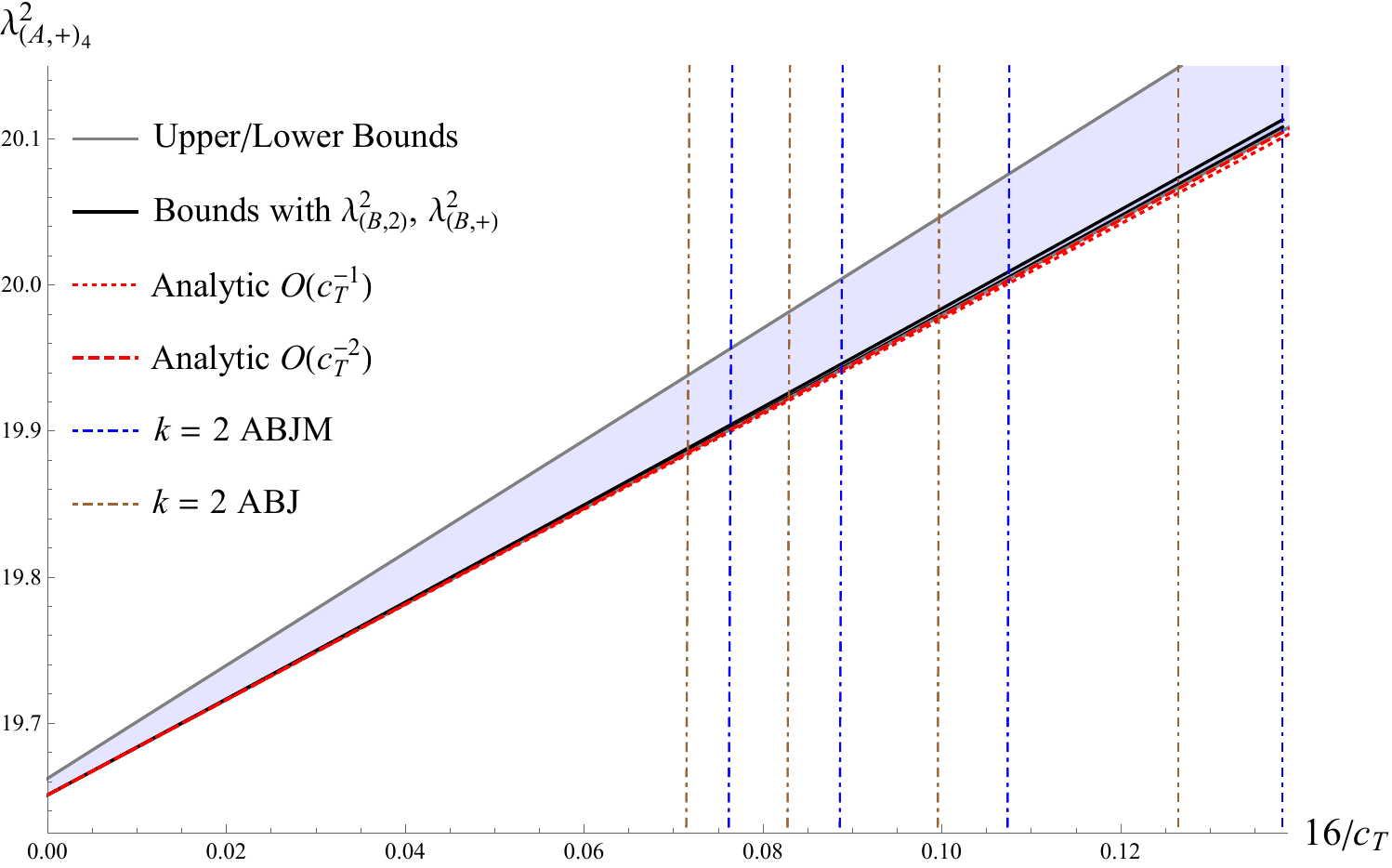}
 \caption{Upper and lower bounds on the $\lambda_{(A,+)_\ell}^2$ OPE coefficient for $\ell=2$ (top) and $\ell=4$ (bottom) in terms of the stress-tensor coefficient $c_T$ in the large $c_T$ regime. The black solid lines are with the all orders in $1/c_T$ values of $\lambda_{(B,+)}^2$ and $\lambda_{(B,2)}^2$ for $k=2$ ABJ(M) inputed into the bootstrap, while the gray solid lines are without any assumptions, and note that the allowed region for the black bounds is much smaller than the gray bounds. The red dotted line denotes the large $c_T$ expansion to order tree level supergravity $O(c_T^{-1})$, which is independent of $k$. The red dashed line also includes the 1-loop $R|R$ correction at order $O(c_T^{-2})$, which depends on the value $k=2$, and improves the saturation of the lower bound relative to $O(c_T^{-1})$. The blue and brown vertical lines denote the values of $c_T$ for various known $k=2$ ABJM and ABJ theories, respectively, which are summarized in Table \ref{cTValues}. These plots were made with $\Lambda=83$.}
\label{Apfig}
\end{center}
\end{figure}

\begin{figure}[]
\begin{center}
   \includegraphics[width=0.85\textwidth]{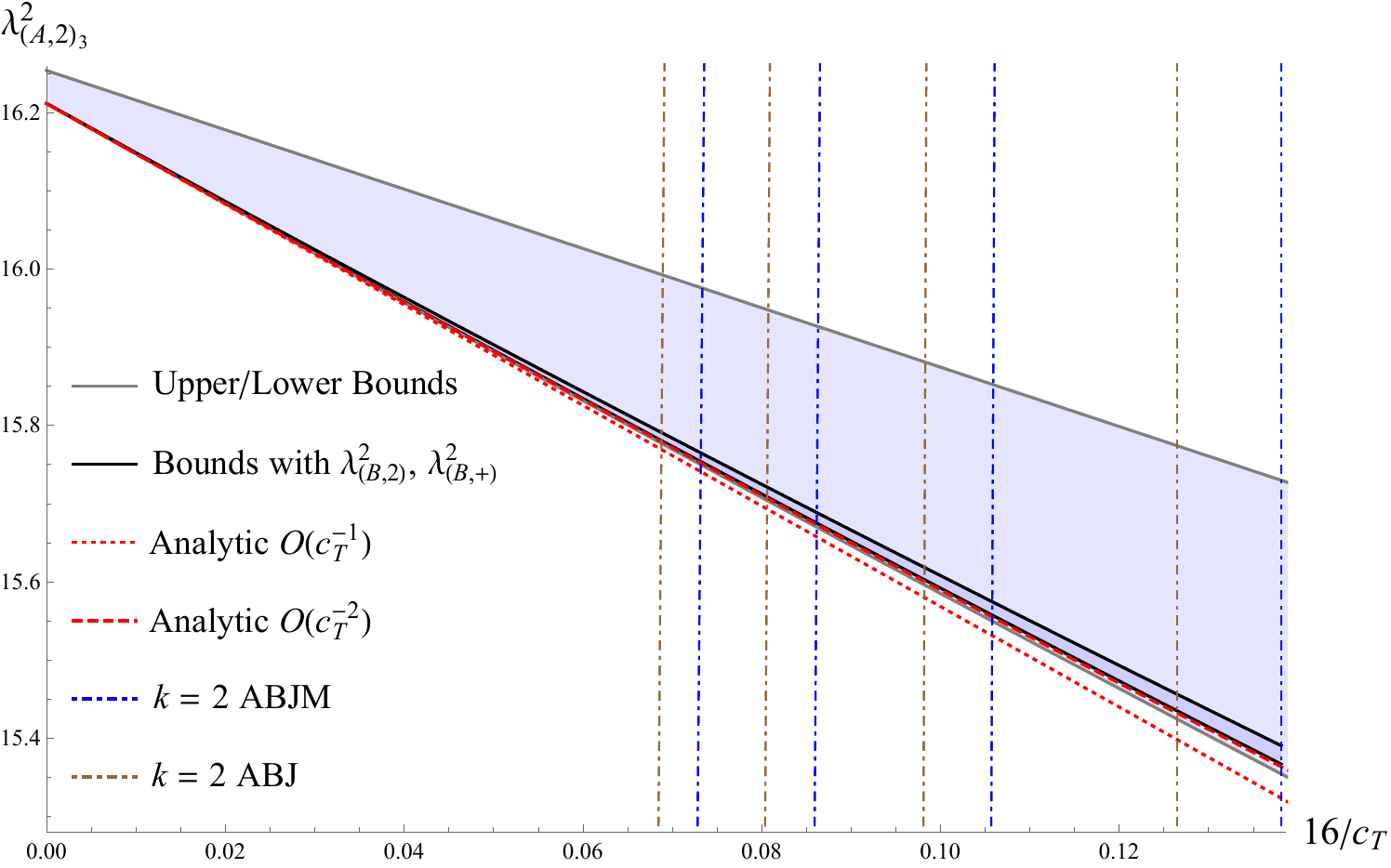}
     \includegraphics[width=0.85\textwidth]{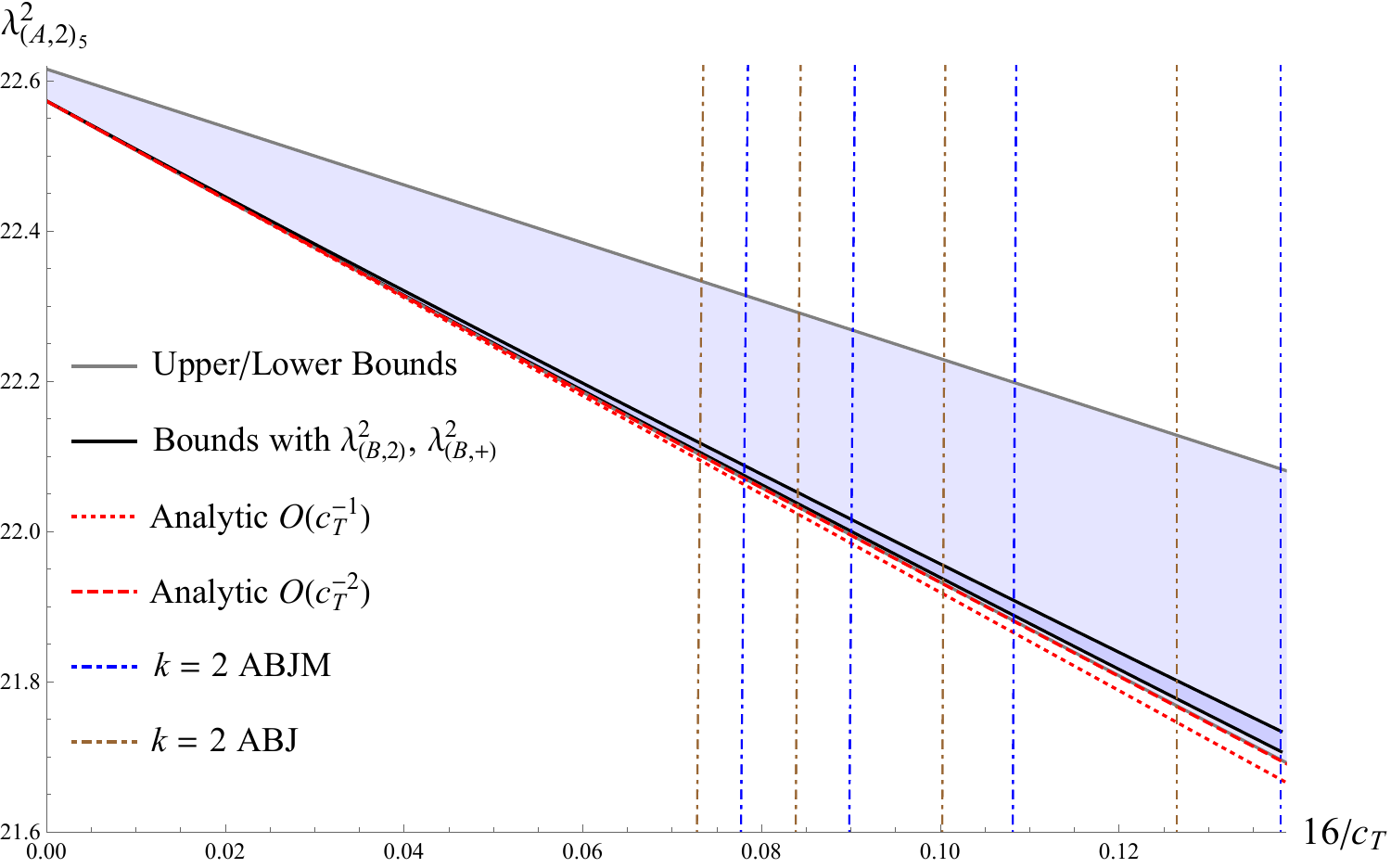}
  \caption{Upper and lower bounds on the $\lambda_{(A,2)_\ell}^2$ OPE coefficient for $\ell=3$ (top) and $\ell=5$ (bottom) in terms of the stress-tensor coefficient $c_T$ in the large $c_T$ regime. The black solid lines are with the all orders in $1/c_T$ values of $\lambda_{(B,+)}^2$ and $\lambda_{(B,2)}^2$ for $k=2$ ABJ(M) inputed into the bootstrap, while the gray solid lines are without any assumptions, and note that the allowed region for the black bounds is much smaller than the gray bounds. The red dotted line denotes the large $c_T$ expansion to order tree level supergravity $O(c_T^{-1})$, which is independent of $k$. The red dashed line also includes the 1-loop $R|R$ correction at order $O(c_T^{-2})$, which depends on the value $k=2$, and improves the saturation of the lower bound relative to $O(c_T^{-1})$. The blue and brown vertical lines denote the values of $c_T$ for various known $k=2$ ABJM and ABJ theories, respectively, which are summarized in Table \ref{cTValues}. These plots were made with $\Lambda=83$.}
\label{A2fig}
\end{center}
\end{figure}

These observations motivate us to focus on comparing to numerical bootstrap for $\lambda^2_{(A,2)_\ell}$ for $\ell>1$ and $\lambda^2_{(A,+)_\ell}$ for $\ell>0$ up to $O(c_T^{-2})$, where we can be reasonably confident that the asymptotic expansion is well converged for a moderately large range of $c_T$. In Figures \ref{Apfig} and \ref{A2fig} we compare the large $c_T$ expansion of this CFT data to non-perturbative lower bounds from the numerical bootstrap in the large $c_T$ regime, which includes many physical examples of $k=2$ ABJ(M) theories as summarized in Table \ref{cTValues}. As discussed above, we show bounds where we inputed the values of $\lambda^2_{(B,2)}$ and $\lambda^2_{(B,+)}$ using their all orders in $1/c_T$ expressions, as well as bounds with no assumptions. The discrepancy between each type of bound is small, and we observe that both are well approximated by the $O(c_T^{-2})$ analytic expressions for the entire range of $c_T$ that we looked at, and that the discrepancy between the two kinds of bounds is smaller than the improvement of the $c_T^{-2}$ term relative to the $O(c_T^{-1})$ approximation. For the more protected $\frac14$-BPS multiplet $(A,+)_\ell$, the correction to tree level supergravity is quite small, so it is somewhat harder to see the 1-loop correction, but for the less protected $\frac18$-BPS multiplet $(A,2)_\ell$ we can very clearly see the improvement from the 1-loop correction to tree level supergravity. We also show upper bounds, which after imposing $\lambda^2_{(B,2)}$ and $\lambda^2_{(B,+)}$ become very close the lower bounds, which suggests that the theory is almost completely fixed.

\section{Conclusion}
\label{conc}

There are three main results of this work. Firstly, we computed the 1-loop terms $R|R$, $R|R^4$, and $R^4|R^4$ for $k=2$ ABJ(M) theory up to contact term ambiguities, and checked that they match the relevant terms in the 11d M-theory S-matrix in the flat space limit. Secondly, we fixed the contact terms for $R|R$ and $R|R^4$ by combining two constraints from supersymmetric localization with a conjectured analytic continuation of the Lorentzian inversion formula, where localization confirms the inversion results for $R|R$ and provides a nontrivial check for $R|R^4$. Finally, we found that the $R|R$, i.e. $c_T^{-2}$, correction to semishort CFT data saturates the numerical bootstrap bounds for $k=2$ ABJ(M) in the large $c_T$ regime.

One could try to perform the same analytic continuation of the Lorentzian inversion formula for the other maximally supersymmetric holographic CFTs that have been studied at 1-loop: 4d $SU(N)$ $\mathcal{N}=4$ SYM dual to Type IIB string theory on $AdS_5\times S^5$, and 6d $(2,0)$ theory dual to M-theory on $AdS_7\times S^4$ for the $A_{N-1}$ theories and $AdS_7\times S^4/\mathbb{Z}_2$ for the $D_{N}$ theories. In the 4d case, general spin formulae were found for the $R|R$ correction in \cite{Aprile:2017bgs} and the $R|R^4$ correction in \cite{Alday:2018pdi}. The $R|R$ formula can be analytically continued to spin zero, which is the spin that is affected by the 4d analogue of the $B_4^{R|R}M^4$ contact term in \eqref{M2222}, but it was shown using supersymmetric localization in \cite{Chester:2019pvm} that $B_4$ is nonzero, unlike what we found here in 3d. The $R|R^4$ formula has explicit poles for spins $0,2,4$, which are the spins affected by the 4d analogue of the $O(c_T^{-\frac83})$ contact terms in \eqref{M2222}, so we cannot analytically continue as we did in here in 3d. In 6d, one can check the Lorentzian inversion formula results in \cite{Alday:2018pdi} can be analytically continued to all CFT data for both $R|R$ and $R|R^4$, which is even better than what we observed in 3d where $R|R^4$ could not be continued to CFT data affected by the $B_4^{R|R}M^4$ contact term. Unlike 3d, however, in 6d we have no localization results to check if this conjectured analytic continuation is correct. 

The similarity of ABJ(M) and 6d $(2,0)$ in contrast to $\mathcal{N}=4$ SYM suggests that the analytically continued Lorentzian inversion formula can be applied to holographic CFTs dual to 11d M-theory, but not those dual to 10d string theory. One could try to justify this by observing that contact terms must always correspond to even powers of Planck length, so they can affect 1-loop terms for 10d duals that are also even in Planck length, but they cannot effect 1-loop terms for 11d duals that are odd in Plank length. On the other hand, recall that we did not compute $M^{R|R}$ and $M^{R|R^4}$ using explicit Witten diagrams, so the difference between a 1-loop diagram and a contact diagram at the same order in $c_T$ is not entirely clear from our approach. One could tentatively define the 1-loop diagram $M^{R|R}$ as whatever gives a formula for CFT data that is analytic in spin for all spins, as we did in practice, and then define $B_4^{R|R}M^4$ as a putative contact term, which we found was zero. Unfortunately, this description would not make sense for $M^{R|R^4}$, where there is no general spin formula that converges for the spins that contribute to $B_4^{R|R^4}M^4$, so it is hard to distinguish between 1-loop diagrams and contact diagrams at the same order $c_T^{-\frac83}$.

It would be nice to check if the analytically continued inversion formula at order $c_T^{-\frac{10}{3}}$ can be used to fix all the contact term ambiguities, perhaps when combined with supersymmetric localization, and if the conjectured results could be checked with localization. In this work we already computed the $R^4|R^4$ term that contributes at order $c_T^{-\frac{10}{3}}$, but we did not yet compute the $R|D^6R^4$ term that also contributes at this order. To compute this term we would need to know $\langle 22pp\rangle$ for all even $p$ at tree level $D^6R^4$, i.e. order $c_T^{-\frac73}$. Currently we only know this for $p=2$, where we could fix the three coefficients in \eqref{M2222} using the two localization constraints from \cite{Chester:2018aca,Agmon:2017xes} and \cite{Binder:2018yvd,Binder:2019mpb} as well as comparison to the known term \eqref{SGtoR4} in the M-theory S-matrix in the flat space limit. For $p>2$, we still have the flat space limit constraint, and it is possible that we could generalize to $p>2$ the localization constraint in \cite{Chester:2018aca,Agmon:2017xes}, which is just the value of short operator OPE coefficients that could in principle be computed from a $k=2$ generalization of the 1d theory \cite{Dedushenko:2016jxl}, which is currently only known for $k=1$ ABJM. On the other hand, the second localization constraint \cite{Binder:2018yvd,Binder:2019mpb} is specific to $p=2$, and also the $p>2$ correlator has more coefficients that need fixing than $p=2$ because crossing constraints are weaker for mixed correlators. Perhaps the $p>2$ term could be computed from the known $p=2$ case, as both in principle arise from dimensional reduction of the same $D^6R^4$ term in the flat space 11d action effective action for M-theory.

 It would be also be very interesting if more terms in the analytic large $c_T$ expansion of CFT data could be found to saturate the numerical bootstrap bounds, beyond the $c_T^{-2}$ terms for semishort OPE coefficients that we considered in this work. In particular, if we could successfully match the $c_T^{-\frac83}$ terms computed in this work, then this would allow us to read off the $D^8R^4$ term at the lower order $c_T^{-\frac{23}{9}}$, which could then be used to fix the corresponding unknown term in the 11d M-theory S-matrix as outlined in \cite{Chester:2018aca}. The coefficients in the asymptotic large $c_T$ expansion start to grow drastically starting at $c_T^{-\frac83}$ even for the best behaved case of semishort OPE coefficients, so to accurately compare them to the numerical bootstrap we must look at very large $c_T$, which requires a very accurate numerical bootstrap. In this work we pushed the current numerical bootstrap to very high accuracy, as parameterized by $\Lambda=83$, which is twice the value used in the previous study \cite{Agmon:2019imm}. We found that the current numerical bootstrap has already started to converge, so pushing to higher $\Lambda$ will likely not improve matters. On the other hand, we observed in this work that inputting the short OPE coefficients noticeably improved the bounds, such that the lower bounds actually become very close to the upper bounds for the regime of $c_T$ that we studied, so it is likely that imposing other exact quantities such as the integrated constraint in \cite{Binder:2018yvd,Binder:2019mpb} will further improve the accuracy of the bounds, and maybe even fix the theory by having upper and lower bounds approximately coincide. It is also likely that a third localization constraint can be computed by considering derivatives of the squashed sphere free energy, which was computed to all orders in $1/c_T$ in \cite{Chester:2021gdw} using the localization results in \cite{Hama:2011ea,Imamura:2011wg}. This additional constraint could both improve the numerical bootstrap, as well as allow us to analytically fix another of the coefficients in \eqref{M2222} for $D^8R^4$, so there would only be a single remaining unfixed coefficient to be fixed from numerics. 

 In this paper we focused on $k=2$ ABJ(M) theory, because it  is technically more difficult to compute the $\langle 22pp\rangle$ data for odd $p$ that is needed to compute $k=1$ ABJM at 1-loop. In particular, while the $\langle 22pp\rangle$ tree level supergravity for even $p$ could be written in terms of a finite number of $\bar{D}_{r_1,r_2,r_3,r_4}(U,V)$, for odd $p$ we require an infinite number. If we could compute $k=1$ ABJM at 1-loop, then we could perform another nontrivial check of our conjectured analytic continuation of the Lorentzian inversion formula, and also try to compare to numerical bootstrap bounds. The numerical bootstrap for $k=1$ could be more accurate than the $k=2$ case in this work, because for $k=1$ we could use the mixed correlator setup in \cite{Agmon:2019imm} that only applies to the $k=1$ theory, and includes inputs from more short OPE coefficients that can be computed to all orders in $1/c_T$ using the localization results of \cite{Gaiotto:2020vqj}. We look forward to reporting on the $k=1$ case in the future.

Finally, it would be interesting to generalize our 1-loop derivation to $\mathcal{N}=6$ ABJ(M) with $k>2$, which at small $k$ is dual to M-theory on $AdS_4\times S^7/\mathbb{Z}_k$ and at large $k$ is dual to Type IIA string theory on $AdS_4\times\mathbb{CP}^3$. The single trace operators in this case can be either $\frac12$ or a $\frac13$ BPS \cite{Dolan:2008vc}, so the unmixing problem would be more complicated. In particular, one would need to derive the superblock expansion of correlators of the various single operator operators to extract the GFFT and tree level data needed to compute 1-loop. So far, the superblock expansion was only computed for the stress tensor multiplet correlator \cite{Binder:2020ckj}, and it seems difficult to generalize that derivation to arbitrary correlators of other $\frac13$-BPS operators. Instead, it might be easier to compute the relevant CFT data by directly imposing the superconformal Ward identity on 3-point functions, rather than expanding 4-point correlators in superblocks. If one could compute the 1-loop term for general $k$, then in the flat space limit one could interpolate between an 11d box diagram at small $k$ to a 10d box diagram at large $k$, just as the tree level $R^4$ term computed in \cite{Binder:2019mpb} was found to interpolate between M-theory and string theory.

\newpage
\section*{Acknowledgments} 

We thank Ofer Aharony, Silvu Pufu, Walter Landry, and Xinan Zhou for useful conversations, Silviu Pufu and Xinan Zhou for collaboration at an early stage of the collaboration, and Ofer Aharony for reading through the manuscript. The work of LFA is supported by the European Research Council (ERC) under the European Union's Horizon
2020 research and innovation programme (grant agreement No 787185). SMC is supported by the Zuckerman STEM Leadership Fellowship. HR acknowledges the support from the PBC postdoctoral fellowship program as well as the Israel Science Foundation center for excellence grant (grant number 1989/14) and by the Minerva foundation with funding from the Federal German Ministry for Education and Research. The authors would like to acknowledge the use of the University of Oxford Advanced Research Computing (ARC) facility in carrying out this work.(http://dx.doi.org/10.5281/zenodo.22558)
\appendix

\section{Conformal blocks and $\bar{D}_{r_1,r_2,r_3,r_4}(U,V)$}
\label{block3dApp}

The $SO(3,2)$ conformal blocks $G_{ \Delta , \ell}(U, V)$ which appear in the decomposition of four-point function of identical scalar operators are eigenfunctions of the quadratic Casimir of $SO(3,2)$. These eigenfunctions are not known in closed form but there exists various expansions. In what follows we present two such expansions which we used heavily for computations. 

The first is the lightcone expansion which is a power series expansion around $U=0$
\begin{equation}
G_{ \Delta , \ell}(U, V)= \sum_{n=0}^{\infty} U^{ (\Delta-\ell)/ 2+n} g^{[n]}_{ \Delta ,\ell}(1-V)
\end{equation}
where $g^{[n]}_{\Delta ,\ell}(1-V)$ are the lightcone blocks that depend on the dimension $ \Delta $ and spin $\ell$ of the exchanged operator. In our normalization, these lightcone blocks take the following form \cite{Li:2019cwm}
\begin{align}
g^{[n]}_{\Delta,\ell}(1-V) = \frac{1}{4^{\Delta}}\frac{\left(\frac{1}{2}\right)_\ell}{\ell !} \sum_{m=0,2,4,...2n} A_{n, m} f_{n, m}(1-V)
\end{align}
where the basis functions $f_{n, m}$ are ${}_2F_1$ hypergeometric functions given by
\begin{equation}
f_{n,m}(1-V)=(1-V)^{\ell-m}{ }_{2} F_{1}\left[\begin{array}{c}
 (\Delta+\ell)/2+n-m,   (\Delta+\ell)/2+n-m \\
2( (\Delta+\ell)/2+n-m)
\end{array} ; 1-V\right]
\end{equation}
and the expansion coefficients $A_{n,m}$ are 
\begin{equation}
\begin{aligned}
A_{n,m}=& \sum_{m_{1}, m_{2}=0}^{n_{1}}(-1)^{m+m_{1}+1} 4^{m_{1}+m_{2}} \frac{(-\ell)_{m}\left(-n_{1}\right)_{m_{1}+m_{2}}\left(n-n_{1}+1 / 2\right)_{m_{1}}}{m ! m_{1} ! m_{2} !\left(n-m+m_{1}\right) !} \\
& \times \frac{(  \Delta -1)_{2 n-m}(3 / 2-\Delta)_{m-n-m_{1}-m_{2}}(2- \Delta+\ell)_{2\left(n_{1}-m_{2}\right)-m}}{(  (\Delta-\ell)/2 +2 \ell-m-1)_{2 n-m}(  (\Delta+\ell)/2 +2)_{2\left(n+m_{1}-n_{1}\right)-m}} \\
& \times \frac{\left(-1 / 2-\ell-n+m-m_{1}+m_{2}\right)\left(n_{2}-\ell\right)_{m_{2}}}{(1 / 2-\ell)_{-n+m+m_{2}}\left(1 / 2+\ell-m_{2}\right)_{n-m+m_{1}+m_{2}+1}} \\
& \times \left[\left(\frac{ \Delta+\ell }{2}\right)_{n-m+m_{1}} \left(\frac{\Delta-\ell-1}{2}\right)_{m_{2}} \right]^4 
\end{aligned}
\end{equation}
in which $n_1, n_2$ are defined as
\begin{equation}
n_{1}=\lfloor m / 2\rfloor, \quad n_{2}=\lfloor(m+1) / 2\rfloor~.
\end{equation}
Setting $n=0$ in the above formulas gives us the leading lightcone block in \eqref{lightconeBlock}. This form of the conformal blocks was particularly useful for the extraction of the tree-level CFT data and in computing the $U^n \log^2 U$ slices in section \ref{1loopfrom}.

The second convenient representation of the conformal blocks is in terms of expansion in Jack polynomials. In \cite{Dolan:2003hv}, it was shown that in three dimensions the conformal blocks can be written as (in a normalization identical to those of the lightcone blocks)
\begin{align}
G_{\Delta, \ell}(z, \bar{z})=\frac{1}{4^\Delta}\frac{\left(\frac{1}{2}\right)_\ell}{\ell!} \sum_{m,n} r_{m,n}(\Delta,\ell) J_{m,n}^{(\Delta,\ell)}(z,\bar{z})
\end{align}
where the Jack polynomials  $J_{m,n}$ in three dimensions are given in terms of Legendre polynomials as follows
\be
J_{m,n}^{(\Delta,\ell)}(z,\bar{z}) = (z \bar{z})^{\frac{1}{2} (m+n+\Delta )} P_{m-n+\ell }\left(\frac{z+\bar{z}}{\sqrt{4 z \bar{z}}}\right)~.
\ee
The expansion coefficients $r_{m,n}(\Delta,\ell)$ are given by
\begin{align}
&r_{m,n}(\Delta,\ell)=-\pi  4^{\Delta -2} \Gamma \left(\Delta -\frac{1}{2}\right) \frac{\Gamma (\ell +1)}{\Gamma (\Delta -1) } (2 m-2 n+2 \ell +1) \csc (\pi  (m+\ell ))  \no\\[5pt]
& \times \frac{\Gamma \left(\frac{\Delta -\ell }{2}\right) \Gamma \left(\frac{1}{2} (\ell +\Delta +1)\right) \Gamma \left(\frac{1}{2} (2 m+\ell +\Delta )\right)^2 \Gamma \left(\frac{1}{2} (2 n-\ell +\Delta -1)\right)^2}{m!  n! \Gamma \left(\ell +\frac{1}{2}\right) \Gamma \left(m+\ell +\frac{3}{2}\right) \Gamma \left(\frac{1}{2} (-\ell +\Delta -1)\right) \Gamma \left(\frac{\ell +\Delta }{2}\right) \Gamma (m+\ell +\Delta )}\no\\[5pt]
&\times \, _4F_3\left(\frac{1}{2},-m+n-\ell ,-\ell ,\Delta -1;1,-m-\ell ,n+\Delta - \ell -\frac{1}{2};1\right)
\end{align}
The poles from the $\csc (\pi  (m+\ell ))$ cancels with the corresponding zeros of ${}_4F_3$ and therefore the above coefficients are well defined in the allowed range of $m$ and $\ell$. This representation of the 3d conformal blocks was useful for computing the coefficients of the various $\mathcal{N}=8$ superconformal blocks which was discussed in section \ref{qqpp}.

Next, we discuss the $\bar D_{r_1,r_2,r_3,r_4}(U,V)$. They can be compactly written in Mellin space as \cite{Rastelli:2017udc}
  \es{dbarM}{
 & \bar D_{r_1,r_2,r_3,r_4}(U,V)=\int\frac{ds\, dt}{(4\pi i)^2} U^{\frac s2}V^{\frac t2}M(s,t)   \Gamma (-\frac{s}{2})  \Gamma (-\frac{t}{2}) \Gamma \left(r_2+\frac{s+t}{2}\right)\\
 & \Gamma \left(\frac{1}{2}
   (-r_1-r_2+r_3+r_4)-\frac{s}{2}\right) \Gamma \left(\frac{1}{2} (r_1-r_2-r_3+r_4)-\frac{t}{2}\right)
   \Gamma \left(\frac{1}{2} (r_1+r_2+r_3-r_4)+\frac{s+t}{2}\right) \,.\\
    }
We would also like the position space expression of their $\log U$ terms for the purpose of extracting the tree-level anomalous dimensions, which are given in \cite{Dolan:2000ut} as\footnote{In the case at hand, the $\bar{D}$ functions involve half-integer arguments so one has to be careful in reading off the $\log U$ pieces by paying attention to the location of the poles of $\Gamma$ functions. However upon inspection it is found that the arguments are such that formula (C.8) of \cite{Dolan:2000ut} is still applicable.}
\begin{align}
& \bar D_{r_1,r_2,r_3,r_4}(U,V)\big|_{\log U}= -e^{\frac{1}{2} i \pi  ( r_1 + r_2 - r_3 - r_4 )}\sum_{m,n=0}^\infty U^m (1-V)^n \times\no\\[10pt]
& \frac{ \Gamma ( r_1 +m)  \Gamma ( r_2 +m+n) \Gamma \left(\frac{1}{2} ( r_1 + r_2 - r_3 + r_4 )+m\right) \Gamma \left(\frac{1}{2} ( r_1 + r_2 + r_3 - r_4 )+m+n\right)}{\Gamma (m+1) \Gamma (n+1) \Gamma ( r_1 + r_2 +2 m+n) \Gamma \left(\frac{1}{2} ( r_1 + r_2 - r_3 - r_4 +2)+m\right)}\,.
\end{align}
The arguments $r_i$ of the $\bar{D}$ functions that appear in the $22pp$ tree-level correlator is such that the overall phase is always real.

\section{Lorentzian inversion at large $c_T$}
\label{Lorentz}

In this Appendix we explain how to extract the CFT-data at one-loop from the double discontinuity (DD) of a 3d CFT using the Lorentzian inversion formula following the similar 4d case in \cite{Alday:2017vkk}. The function that describes the CFT-data is given by
\es{app1}{
c(\Delta,\ell)= {\cal N}(\Delta,\ell)\int_0^1 dz d\bar z \frac{\bar z-z}{(z \bar z)^3} G_{\ell+2,\Delta-2}(U,V) \text{dDisc}[\mathcal{G}(z,\bar z)]\,,
}
where ${\cal N}(\Delta,\ell)$ is some normalization factor that we can fix by comparing to GFFT, and $\mathcal{G}(z,\bar z)$ is the 4-point function with the usual $x$-dependence factored out as in \eqref{4point}. Note that the inversion formula exchanges the roles of dimension and spin in the standard conformal block $G_{\Delta,\ell}(U,V)$.

An intermediate operator of a given twist $t$ will produce a pole whose residue is related to its three-point function. We can see this as follows. Its contribution to the DD in a small $z$  expansion is given by (recall the DD is taken in the $\bar z$ variable around 1)
\begin{equation}
\text{dDisc}[\mathcal{G}(z,\bar z)]= z^{t/2} h(\bar z) + \cdots\,,
\end{equation}
for some function $h(\bar z)$ that is analytic at $\bar z=1$.  On the other hand, the conformal block is expanded at small $z$ as in \eqref{lightBlocksExp} in terms of the leading lightcone block $g_{\Delta,\ell}(\bar z)$ in \eqref{lightconeBlock}, which for $G_{\ell+2,\Delta-2}(U,V)$ in \eqref{app1} gives 
\es{lightApp}{
G_{\ell+2,\Delta-2}(U,V)&=z^{2-(\Delta-\ell)/2}g_{\ell+2,\Delta-2}(\bar z)+\dots\,,\\
}
 where we dropped the superscript on the leading lightcone block for simplicity and at leading order in $z$ we set $V=(1-\bar z)$. Plugging these expansions into the inversion formula we see that the integral over $z$ produces a pole at $\Delta-\ell=t$. If we now expand at large $c_T$, then we will have operators of even integer twist $t$ and spin $\ell$, which are corrected by anomalous dimensions $\gamma_\ell$ that depend on $\ell$ and go to zero as $c_T\to\infty$. The pole at $t$ at large $c_T$ then gives
\begin{equation}
c(\Delta,\ell)=\frac{a_\ell}{\gamma_\ell+t-(\Delta-\ell)} \,,
\end{equation}
where $a_\ell$ denotes the OPE coefficients squared for the operator with approximate twist $t$, and we have assumed that the leading twist operators under consideration are non-degenerate (which will be the case for our application). To expand the poles, it is useful to use the normalization of \cite{Caron-Huot:2017vep}. Our blocks obey
\[
G_{\Delta,\ell}=n_{\Delta,\ell}G^{\rm CH}_{\Delta,\ell}\,,\qquad
n_{\Delta,\ell}=\frac{\Gamma(\ell+1/2)}{4^\Delta\sqrt{\pi}\ell!}\,.
\]
Thus $c^{\rm CH}(\Delta,\ell)=n_{\Delta,\ell}c(\Delta,\ell)$ has physical residues $a^{\rm CH}_\ell=n_{t+\ell+\gamma_\ell,\ell}a_\ell$. Expand $a^{\rm CH}_\ell=a^{\rm CH,(0)}_\ell+c_T^{-1}a^{\rm CH,(1)}_\ell+\cdots$ and $\gamma_\ell=c_T^{-1}\gamma^{(1)}_\ell+\cdots$. At order $c_T^{-2}$, writing $\beta=\Delta+\ell$, we get\footnote{We can get an analogous formula also at other 1-loop orders such as $c_T^{-\frac83}$.}
\begin{equation}
c^{\rm CH}(\Delta,\ell)=-\frac{R^{\rm CH}_0(\beta)}{((\Delta-\ell-t)/2)^3}-\frac{R^{\rm CH}_1(\beta)}{((\Delta-\ell-t)/2)^2}-\frac{R^{\rm CH}_2(\beta)}{(\Delta-\ell-t)/2}+\cdots\,.
\end{equation}
Expanding the $\beta$ dependence at fixed $\ell$ and matching the poles gives
\es{app2}{
a^{\rm CH,(0)}_\ell(\gamma^{(1)}_\ell)^2&=8R^{\rm CH}_0\,,\\
a^{\rm CH,(0)}_\ell\gamma^{(2)}_\ell+a^{\rm CH,(1)}_\ell\gamma^{(1)}_\ell&=4R^{\rm CH}_1+8\partial_\beta R^{\rm CH}_0\,,\\
a^{\rm CH,(2)}_\ell&=2R^{\rm CH}_2+4\partial_\beta R^{\rm CH}_1+4\partial_\beta^2 R^{\rm CH}_0\,,
}
where all residues and derivatives are evaluated at $\beta=t+2\ell$. In our normalization, set $R_i=R^{\rm CH}_i/n_{t+\ell,\ell}$. Since $a^{\rm CH,(1)}_\ell=n a^{(1)}_\ell+\gamma^{(1)}_\ell(\partial_\Delta n)a^{(0)}_\ell$, the first two relations become
\begin{align*}
a^{(0)}_\ell(\gamma^{(1)}_\ell)^2&=8R_0\,,\\
a^{(0)}_\ell\gamma^{(2)}_\ell+a^{(1)}_\ell\gamma^{(1)}_\ell
&=4R_1+4\partial_\ell R_0
+4R_0\left.(\partial_\ell-\partial_\Delta)\log n_{\Delta,\ell}\right|_{\Delta=t+\ell}\,.
\end{align*}
Here $\partial_\ell R_0$ is taken at fixed $t$, while the two partial derivatives of $n$ are taken before setting $\Delta=t+\ell$. For integer spin, $\tfrac12(\partial_\ell-\partial_\Delta)\log n_{\Delta,\ell}=A_{2\ell}$. The normalization term therefore adds $A_{2\ell}a^{(0)}_\ell(\gamma^{(1)}_\ell)^2$ to the second relation.
From the inversion integral, these residues are computed as follows. To order $O(c_T^{-2})$ the contribution to the DD from the operators of approximate twist $t$ takes the form
\begin{equation}
\text{dDisc}[G(z,\bar z)] =z^{\frac t2} \left( h_0(\bar z) \log^2 z + h_1(\bar z) \log z+ h_2(\bar z) \right) + \cdots
\end{equation}
for some functions $h_i(\bar z)$. We can then perform the integral over $z$ keeping the relevant poles in a small $z$ expansion to get
\es{app3}{
R^{\rm CH}_0(t+2\ell) &= 2 {\cal N}^{\rm CH}(t+\ell,\ell)\int_0^1 \frac{d \bar z}{\bar z^2}   g_{\ell+2,t+\ell-2}(\bar z) h_0(\bar z) \,,\\
R^{\rm CH}_1(t+2\ell) &= {\cal N}^{\rm CH}(t+\ell,\ell)\int_0^1 \frac{d \bar z}{\bar z^2}   g_{\ell+2,t+\ell-2}(\bar z) h_1(\bar z) \,, \\
R^{\rm CH}_2(t+2\ell) &= {\cal N}^{\rm CH}(t+\ell,\ell) \int_0^1 \frac{d \bar z}{\bar z^2}    g_{\ell+2,t+\ell-2}(\bar z) h_2(\bar z) \,, 
}
where ${\cal N}^{\rm CH}(\Delta,\ell)=n_{\Delta,\ell}{\cal N}(\Delta,\ell)$. Its overall normalization can be fixed by recovering the GFFT coefficients in the same CH convention. This check fixes the normalization on the unperturbed twist but does not remove its dimension dependence. In particular, the large $c_T$ GFFT is proportional to\footnote{For the $\mathcal{N}=8$ theory in the main text, the precise formula for $\langle 2222\rangle$ is $\mathcal{G}_\text{GFFT}(U,V)=1+\sigma^2U+\tau^2\frac{U}{V}$.}
\es{GFFTapp}{
\mathcal{G}_\text{GFFT}(z,\bar z) \propto z\left( \frac{\bar z}{1-\bar z}\right)^p\Big\vert_{p=1}+\text{regular}\,.
}
We can compute the DD as explained in \cite{Caron-Huot:2017vep} by first letting $p$ be a general variable, using
\es{discApp}{
\text{dDisc}[\left(\frac{\bar z}{1-\bar z} \right)^{p}]  =2 \sin^2(\pi p) \left(\frac{\bar z}{1-\bar z} \right)^{p}\,,
}
and then computing $\bar z$ integral in \eqref{app3}, which has a pole at $p=1$ that cancels the zero in \eqref{discApp} to give a finite nonzero answer for $R_0$ in \eqref{app3}. We can then compare to the expected GFFT values to fix the normalization.

\bibliographystyle{JHEP}
\bibliography{3ddraft}

\end{document}